\documentclass[%
 reprint,
 amsmath,amssymb,
 aps,
]{revtex4-2}

\usepackage{graphicx}% Include figure files
\usepackage{dcolumn}% Align table columns on decimal point
\usepackage{bm}% bold math
\usepackage{color}

\usepackage{pdfpages} % \includepdf
\usepackage{pgffor}   % \foreach（TikZ/PGF）
\usepackage{pdfpages}
\makeatletter
\AtBeginDocument{\let\LS@rot\@undefined}
\makeatother

\begin{document}

% \preprint{APS/123-QED}

\title{Wrinkles, rucks, and folds formed in a heavy sheet on a frictional surface}

\author{Keisuke Yoshida$^{1,2}$}
\email{kyosh424@gmail.com}
\author{Hirofumi Wada$^{1}$} 

\affiliation{$^{1}$Department of Physical Sciences, Ritsumeikan University, Kusatsu, Shiga 525-8577, Japan}
\affiliation{$^{2}$Research Organization of Science and Technology, Ritsumeikan University, Kusatsu, Shiga 525-8577, Japan}

\date{\today}

\begin{abstract}
Soft elastic sheets resting on rigid surfaces develop wrinkles, rucks, and folds due to the combined influence of elasticity, gravity, and contact interactions. Despite their ubiquity, the principles governing their morphology and transitions remain unclear. We introduce a minimal experiment in which the center of a gravity-loaded sheet is gradually lifted from the supporting plane. This operation generates a clear sequence of shapes: an axisymmetric uplift, a finite number of wrinkles, system-spanning rucks produced by global buckling, and folded states that can arise from ruck collapse upon unloading at larger lifts. 
Combining experiments, finite-element simulations, and F\"{o}ppl–von K\'{a}rm\'{a}n theory, we establish a unified physical picture of this morphology sequence.
In the frictionless case, elasticity and gravity alone govern the response, leading to a universal wrinkling threshold: the wrinkle number is fixed and the onset displacement scales linearly with the sheet thickness.
With interfacial friction, the wrinkled state is described by introducing an additional nondimensional parameter that compares frictional and elastic–gravitational forces.
These results suggest a simple route to programmable sheet morphogenesis via friction and gravity.
\end{abstract}

\maketitle
\section{Introduction}

\begin{figure*}
    \centering	\includegraphics[width=1.\linewidth]{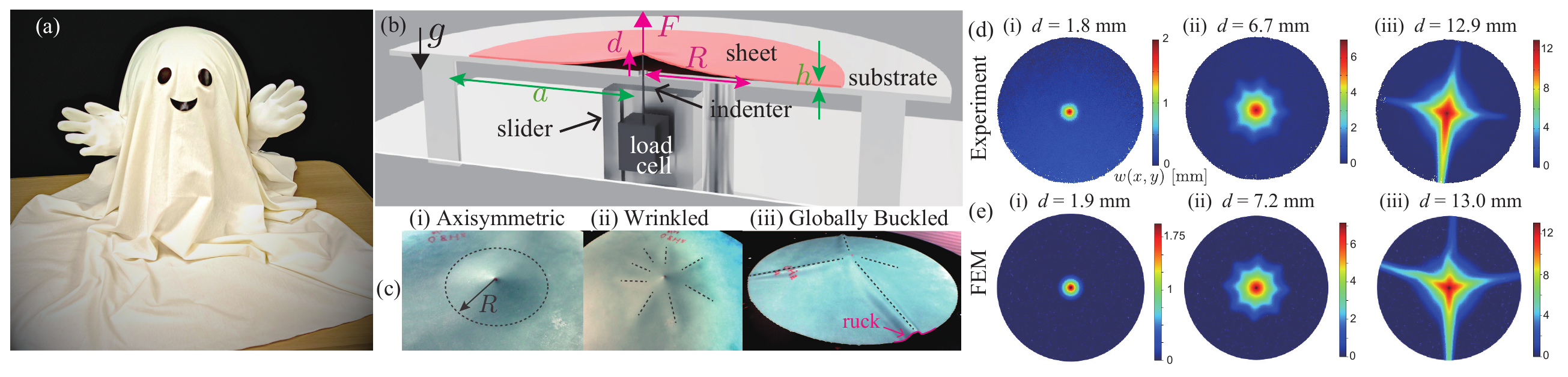}\caption{
    (a) Halloween ghost formed by draping a soft, thin fabric over a convex object.
    (b) Schematic of the experimental setup. A thin elastic sheet of radius $a$ is indented vertically at its center by a distance $d$ from beneath the sheet on a rigid flat substrate. A region with a certain radius $R$ detaches from the substrate. The indentation force $F$ is measured as a function of the indentation displacement $d$ by a load cell attached to the indenter.
    (c) Representative images of three distinct lifted shapes for increasing $d$: (i) axisymmetric, (ii) wrinkled, and (iii) globally buckled.
    (d, e): Color maps of the vertical displacements $z=w(x,y)$ for the indicated values of $d$. 
    (d) The point-cloud data acquired by a 3D scanner in our experiments and (e) those in our FES.
    The parameters used both in experiment and FES are $E=477\, {\rm kPa},  \nu=0.47, \rho=1127\, {\rm kg/m}^3, h = 0.27\, {\rm mm}, a = 104\, {\rm mm}$ and $\mu=0.32.$}\label{Figure01}
\end{figure*}

Soft, thin structures resting on external surfaces frequently develop visually striking wrinkles, rucks, and folds under their own weight. 
A familiar example is the cartoon-like ``ghost'' shape that appears when a soft sheet is draped over a small object on a table~[Fig.~\ref{Figure01}(a)].
Examples of such gravity-induced morphologies span a wide range of systems, from everyday drapery~\cite{cerda2004elements,chen2010non, boedec2021disk, vandeparre2011wrinkling, taffetani2019limitations} to solar sails~\cite{johnston2006analytical, spencer2019solar} and even geophysical plates~\cite{karner1983gravity, mahadevan2010subduction, protiere2017sinking}. 
In thin yet heavy systems, contact interactions can generate unpredictable surface patterns that pose practical concerns~\cite{spencer2019solar}.

Wrinkles may be deliberately introduced through in-plane loading~\cite{cerda2003geometry, chan2008surface, chung2011surface}, but more commonly emerge from the uncontrolled sticking and sliding of thin materials against their surroundings~\cite{yamaguchi2009regular, aoyanagi2010random, chawla2024geometry}.
When a thin sheet interacts with external surfaces, the contact forces and friction significantly complicate the buckling process, producing patterns that are far more diverse than those arising from simple in-plane compression alone~\cite{plaut1999deflections, roman2002postbuckling, stoop2008morphological, liu2013effect, alben2022packing, deboeuf2024yin}. 
These effects are particularly important in gravity-loaded sheets, where the self-weight naturally brings the material into contact with the supporting surface~\cite{mahadevan1996coiling, habibi2007coiling, vella2009statics, kolinski2009shape, sano2017slip}.

Although the mechanics of elastic beams or rods in contact with external surfaces can be analyzed in considerable detail~\cite{vella2009statics, kolinski2009shape, sano2017slip, grandgeorge2022elastic, tani2024soft, curtis2025bridging}, the mechanics of two-dimensional sheets remain far less tractable.
In sheets, the curvature and in-plane stress are coupled through geometric compatibility~\cite{audoly2010elasticity}, making the influence of the contact forces far more complex than in slender one-dimensional systems~\cite{witten2007stress, hure2011wrapping, hure2012stamping, suzanne2022indentation, montalvo2023wrinkling, chawla2024geometry}.
In this context, indentation tests offer an appealing minimal model with a simple axisymmetric geometry. They have been widely used to probe thin films--from biological membranes~\cite{krieg2019atomic} to nanoscale sheets~\cite{akinwande2017review}--and readily reveal symmetry breaking into wrinkles or folds~\cite{chopin2008liquid, holmes2010draping, vella2018regimes, suzanne2022indentation, montalvo2023wrinkling}.

To investigate the mechanism by which gravity and contact interactions shape the morphology of heavy sheets, we introduce an indentation experiment in which an elastic sheet resting on a rigid substrate is lifted gradually from its center [Fig.~\ref{Figure01} (b)]. 
This operation produces a sequence of patterns [Fig.~\ref{Figure01} (c--e)]: initial axisymmetric uplift, a finite number of wrinkles, and then system-spanning rucks produced by buckling at the periphery (\textit{global buckling}). Folded states can also arise upon unloading after large lifts~[Fig.~\ref{Figure13} and Supplemental Material (SM) movies].
Using a combination of tabletop experiments, finite-element simulations (FES), and F\"{o}ppl–von K\'{a}rm\'{a}n theory, we characterize the mechanical response across this sequence. In the frictionless case, the wrinkled state shows remarkable universality: the wrinkle number is fixed, and the critical onset displacement scales linearly with the sheet thickness. When friction is present, it changes the hoop stress distribution, and both the wrinkle number and its onset are described by introducing an additional dimensionless parameter measuring the relative importance of frictional and elastic–gravitational forces.

%%% FIGURE 1

The remainder of this paper is organized as follows.
Section II describes experimental setup, including sample fabrication, apparatus, measurement protocols, and the finite-element methodology.
Section III presents the force response and growth of the lifted region, and identifies a characteristic length scale that organizes these observations.
Section IV develops a theoretical framework based on the F\"{o}ppl–von K\'{a}rm\'{a}n equations and compares its predictions with experiments and simulations.
Section V presents the numerical results for the in-plane displacements and stress distributions in the contact region.
Section VI examines the wrinkling transition through additional experiments, FES, and theoretical considerations, including the effects of friction.
Section VII derives a scaling law for the onset of global buckling and validates it against experimental and numerical results.
Section VIII highlights the hysteresis and fold formation observed during unloading.
Finally, section IX summarizes the main findings and discusses future research directions.

\section{Indentation test}

\subsection{Experiments}

In our experiments, we used circular sheets with a uniform thickness $h$ in the range 0.14~mm$\le h\le 4.96$~mm, and radius $a$ cut in the range of 40~mm $\le a\le 118$~mm.
To obtain sheets of uniform thickness, we performed spin-coating with addition-cure silicone rubbers (Elite Double 8 (Zhermack, Italy), HTV-4000 (Engraving Japan), Mold Star 15 SLOW and Ecoflex 00-20 (Smooth-On, USA)).
Their Young's modulus $E$, Poisson's ratio $\nu$, and mass density $\rho$ are summarized in Table~\ref{Table01}.
Talc was sprinkled to the sheet surfaces to prevent adherence.
We also used a commercially available urethane sheet with $h=2.0$~mm, $E=3780$~kPa, and $\nu= 0.29$.
All the $h$ values were measured using a laser displacement sensor (LK-G3000, KEYENCE, Japan). 
$E$ and $\nu$ are measured using cantilever bending tests or tensile tests.

%%% TABLE 1
\begin{table}
 \begin{center}
   \caption{Properties of the elastic sheets fabricated by authors, including: the material, Young's modulus $E$, Poisson's ratio $\nu$, mass density $\rho$, and the coefficient of static friction between the sheet and substrate $\mu$. }
   \label{Table01}
  \begin{tabular}{ccccc} 
 Material & $E$ [kPa] & $\nu$ & $\rho$ [kg/m$^3$] & $\mu$ \\ \hline
    HTV4000 & 756 & 0.34 & 1142 & $0.34 \pm 0.04$ \\
    Mold Star 15 SLOW & 477 & 0.47 & 1127 &  $0.32 \pm 0.01$ \\
    Elite Double 8 & 226 & 0.49 & 1030 &  $0.44 \pm 0.03$ \\
    Ecoflex 00-20 & 44 & 0.43 & 1053 &  $0.56 \pm 0.03$ \\
  \end{tabular}
 \end{center}
\end{table}

We used a stainless-steel substrate with a diameter of 300~mm and a thickness of 1.5~mm, which can be treated as a semi-infinite rigid plane [Fig.~\ref{Figure01}~(b)].
The substrate had a circular hole with a diameter of 11~mm at its center.
The center of the sheet was pushed through this hole using a cylindrical indenter with a length of 30--70~mm and radius $r_{\rm ind}= 0.69-1.70$~mm.
Because the hole was small but larger than the indenter, airflow could be generated among the hole, indenter, and deformed sheet, preventing any vacuum effects during indentation. 
The coefficient of static friction between the talc-coated elastomer and the stainless-steel substrate was measured to be $\mu=0.32-0.56$ [Table~\ref{Table01}] using a slip-angle measurement experiment.
Before placing the sheet, we spread a cationic surfactant (an antistatic agent, MonotaRO, Japan) on the substrate to suppress the static electricity effects.
After evaporation, we placed a sheet and blew it with a hair dryer to remove the fine pre-wrinkles.

The indenter was positioned beneath the substrate as shown in Fig.~\ref{Figure01}~(b).
The vertical motion of the indenter was controlled using a stepping motor (ARM46AC, ORIENTAL MOTOR, Japan).
The indenter moved sufficiently slowly upward (0.1~mm/s), thereby imposing a vertical displacement $d$ on the sheet.
The reaction force $F$ exerted at the center of the sheet was measured using a load cell (LTS-2KA, KYOWA, Japan) attached to the indenter, and recorded as a function of $d$.

To quantify the lifted shape, we obtained point-cloud data $w(r, \theta)$ using a 3D scanner (EinScan-SP, SHINING 3D, China).
The scanning experiment was conducted separately from the force measurements.
We placed the indentation apparatus on a desktop 3D scanner and carefully raised the indenter by 
$\Delta d\approx 1~\textrm{mm}$ using a hand-controlled labjack; a point-cloud dataset was acquired at each step~[Fig.~\ref{Figure01} (d)].
From these data, we extracted the mean lifted radius $R$ for axi-symmetric and wrinkled states.
After global buckling, where axisymmetry was completely lost, we defined the effective radius through the lifted area $S_{\rm lift}$ as $R\equiv \sqrt{S_{\rm lift}/\pi}$.

Our method is a type of indentation test, often referred to as a ``blister test,'' which is traditionally used to measure the strength of adhesion in thin films~\cite{dannenberg1961measurement}. 
Several studies have examined pattern formation in adhered films (that can still slide laterally)~\cite{chopin2008liquid, dai2018interface, dai2020radial}.
Although motivated by similar considerations, our system differs in that gravity, rather than adhesion, prevents the lifting, and our focus is on the influence of dry friction~\cite{popov2010contact}.

As $h$ becomes significantly larger than the indenter radius $r_{\rm ind}$, the local contact near the indenter is expected to approach that predicted by punch-indentation theory for a Boussinesq-type problem~\cite{johnson1987contact}.
Because the sheets used in our experiments were sufficiently thin, we did not analyze the detailed 3D contact deformation; instead, we focused on the overall shape evolution of the sheet.

\subsection{Finite-element simulations (FES)}

To complement our experimental results, we performed FES using the commercial package Abaqus (Dassault Syst\`{e}mes, France). 
A linear elastic circular sheet was modeled using quadrilateral linear shell elements with reduced integration and finite membrane strain (S4R). 
The element size $\Delta x$ was chosen to be sufficiently small to resolve azimuthal variations such as wrinkle wavelengths, ensuring $\lambda \gg \Delta x$.
The stainless-steel substrate was modeled as a rigid shell, and normal and tangential contact interactions were included.

To prevent the undesired initial penetration, the sheet was initially placed slightly above the substrate and then we dropped onto it without friction.
After equilibration, Coulomb friction was introduced, and the central circular region of radius $r_{\rm ind}$ was raised at 0.1~mm/s, with $r_{\rm ind}/a\lesssim 0.01$.
Both the dropping and indentation steps were computed using an implicit dynamic analysis (ABAQUS step type: Dynamic, Implicit).
Throughout the simulation, the kinetic-to-strain energy ratio was typically $\sim 10^{-7}-10^{-5}$, independent of $\mu$.
Even in frictionless runs exhibiting global buckling, it peaked at only $\sim10^{-2}$, confirming effectively quasi-static behavior.
The geometric and material parameters were almost the same as those used in the experiments, while $\mu$ was varied over a broader-than-typical range, $0 \le \mu \le 2.0$.

When we focused on large displacements ($d/h>10$), $R$ was obtained from the lifted area $S_{\rm lift}$ using $R= \sqrt{S_{\rm lift}/\pi}$, as in the experiments.
Note that the definition of $R$ is valid when the pattern exhibits the axisymmetry and quasi-axisymmetry such as in wrinkled one; however, in the case of globally buckled morphologies, the assumed symmetry is generally broken and $R$ merely serves as a guide to the size of the lifted region. 
Still, by using the common definition of $R$, it is informative for quantifying the magnitude of the lifted area in the different regimes shown in Fig.~\ref{Figure02}.
For smaller deformations ($d/h\lesssim 10$), the lifted area became difficult to identify accurately because the out-of-plane displacement was very small. 
In this regime, we instead extracted $R$ from the azimuthally averaged profile $\overline{w}(r)\equiv (2\pi)^{-1} \oint  w(r, \theta) d\theta$.

To investigate wrinkling instability, we introduced a small random vertical imperfection at the sheet nodes ($|\delta w|/h<1\%$) to initiate symmetry breaking.
Simulations performed with and without this imposed imperfection showed nearly identical force--displacement curves and wrinkle onsets, suggesting that the intrinsic numerical and geometrical imperfections already present in the model were likely sufficient to trigger symmetry-breaking.

\section{Typical morphology and force response}

\begin{figure}[h]
    \centering	\includegraphics[width=1.\linewidth]{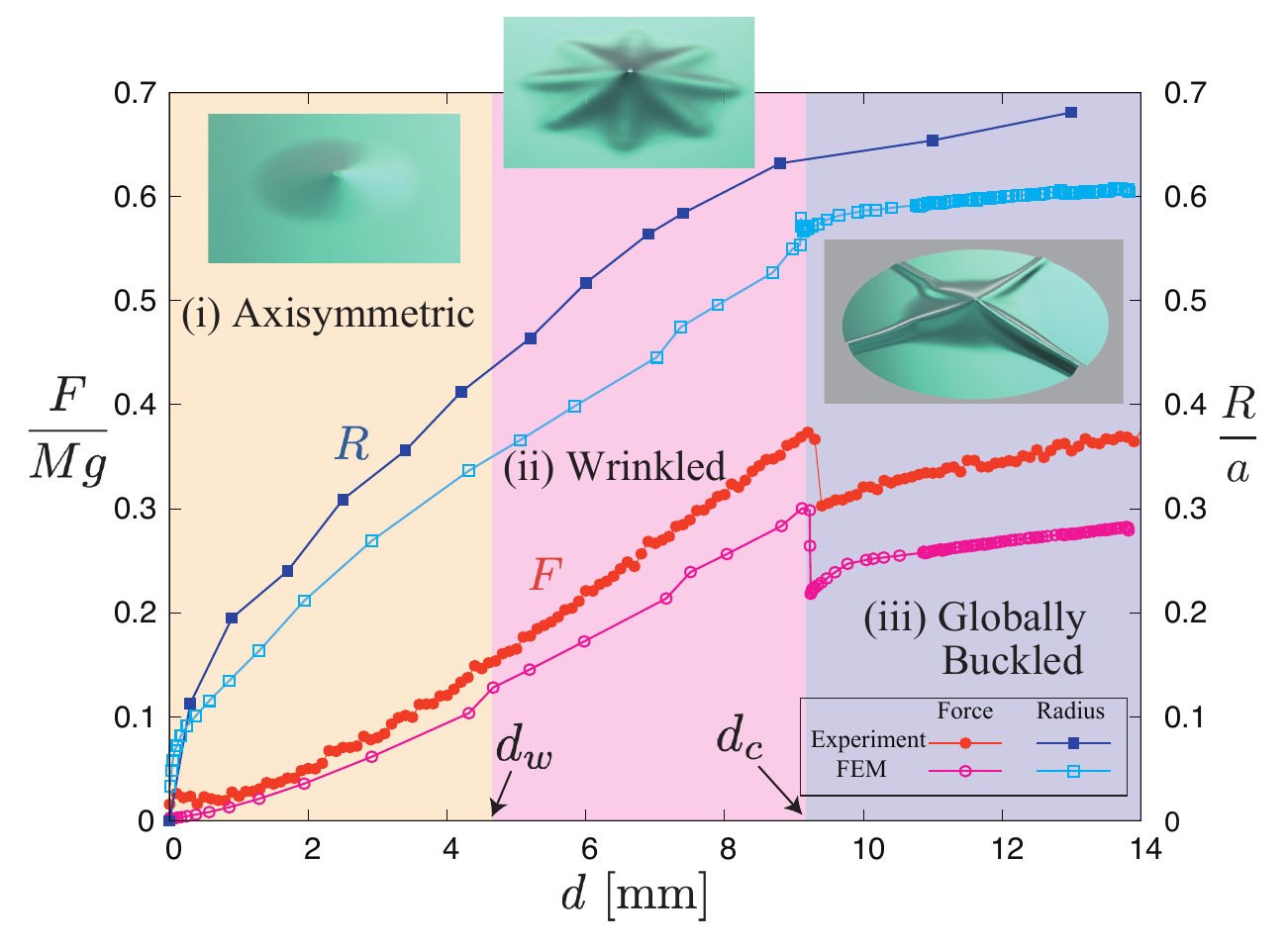} 	\caption{
    Lifting force and lifted radius vs. indentation height.
    The left axis and red points indicate the lifting force $F$, normalized by the total weight of the sheet $Mg$.
    The lifted radius $R$, normalized by the sheet’s full radius $a$, is plotted on the right axis (blue points).
    Filled and open symbols represent data obtained from experiments and FES, respectively.
    The data are taken from the experiment and simulation presented in Fig.~\ref{Figure01}.    
    Inset figures show lifted shapes of the sheet obtained from our FES.
    Axisymmetry breaks at $d=d_w$, where $m$-fold wrinkles emerge ($m$=8 and $d_w\approx 4.9$~mm).
    The critical displacement for global buckling is 
    $d_c\approx 9.2$~mm.
 }\label{Figure02}
\end{figure}

\begin{figure}[]
    \centering	\includegraphics[width=1.\linewidth]{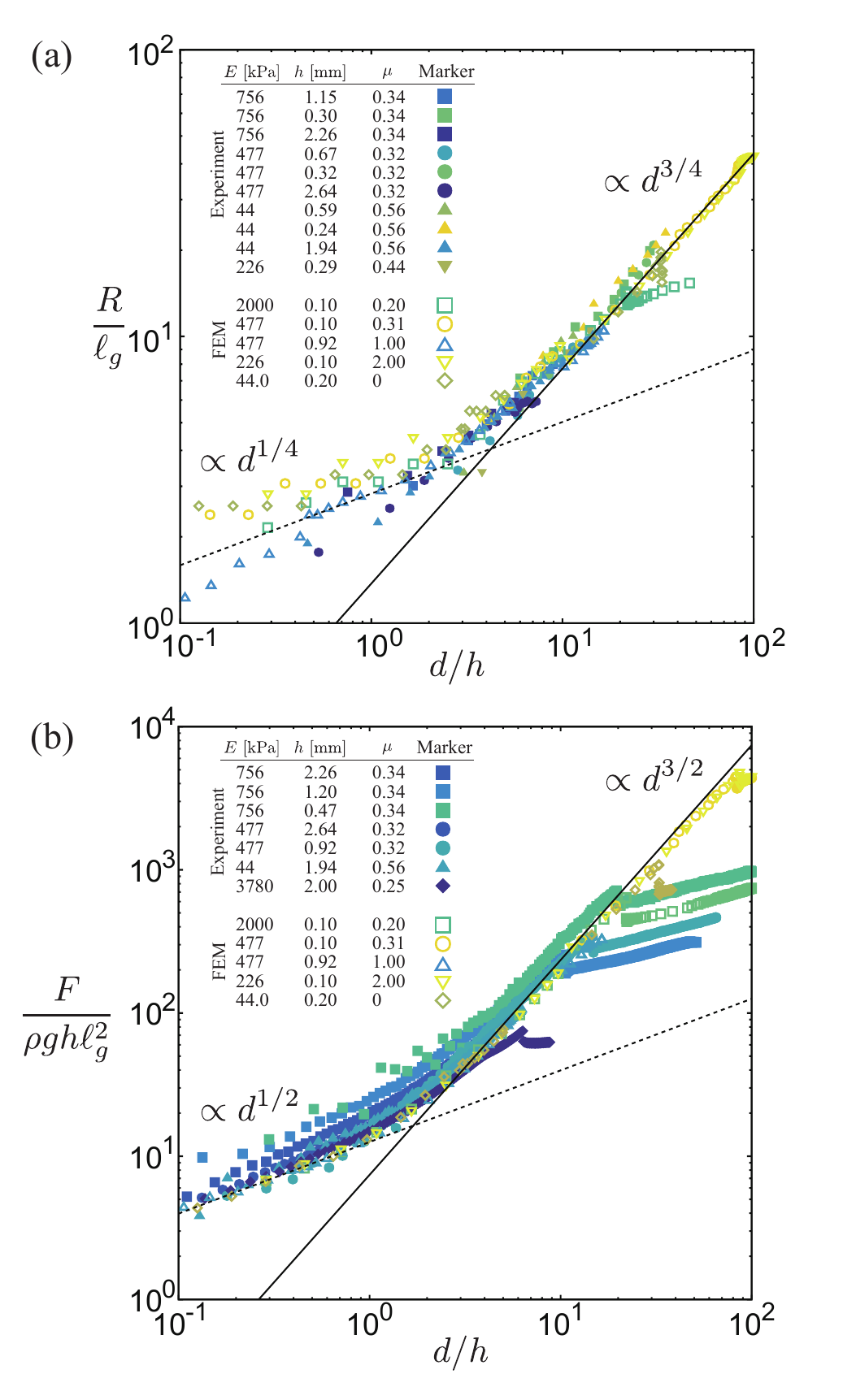} 	\caption{
Dimensionless (a) lifted radius $R/\ell_g$ and (b) lifting force $F/(\rho g h \ell_g^2)$ plotted against normalized central displacement $d/h$.
The characteristic length $\ell_g$ is defined in Eq.~(\ref{sec03:def:ellg}).
Filled and open symbols denotes experimental and FES data, respectively.
The scaling behaviors predicted by Eqs.~(\ref{sec03:eq:R}) and~(\ref{sec03:eq:F}) are confirmed in both the shallow ($d\ll h$) and large indentation ($d\gg h$) regimes, as shown by the dotted and solid lines with the coefficient in Eqs.~(\ref{sec04A:sol:kb}, \ref{sec03:eq:Fd_bend_w_pre}, \ref{sec04B:sol:cs}, and \ref{sec04B:sol:ks}).
 }\label{Figure03}
\end{figure}

Figure~\ref{Figure02} shows the typical force--displacement and lifted radius--displacement curves from both the experiments and FES.
The sheet material was Mold Star 15 Slow, with thickness $h=0.27$ mm and radius $a=104$ mm [material properties in Table~\ref{Table01}].
We integrate some results for the different parameters into Figure~\ref{Figure03}, which shows that both $F$ and $R$ exhibit power law behaviors with respect to $d$ when plotted on a log-log scale. It also appears that the exponents change when $d/h \ll 1$ and $d/h \gg 1$, which will be rationalized by the scaling argument given below.

Although axisymmetry is lost and wrinkles appear at $d=d_w$ (Fig.~\ref{Figure02}), the $F(d)$ and $R(d)$ curves remain monotonic and are unaffected by symmetry breaking.
By contrast, at the onset of global buckling ($d=d_c$), the force exhibits a discontinuous drop, and $R(d)$ curve shows a sharp change in the slope.
After global buckling, $F$ and $R$ increase more gradually.

%%% FIGURE 2

%%% FIGURE 3

We now focus on the power-law behavior of $F$ and $R$ in the axisymmetric and wrinkled phases.
For a lifted region of radius $R$ and height $d$, the typical strain and curvature scales are $\epsilon \sim (d/R)^2$ and $K\sim d/R^2$.
The corresponding stretching and bending energies are as follows:
$\mathcal{E}_s\sim Eh \int_{S_{\rm lift}} \epsilon^2 dS_{\rm lift} \sim Eh{d^4}/{R^2}$ and $\mathcal{E}_b \sim B \int_{S_{\rm lift}} K^2 dS_{\rm lift} \sim Eh^3{d^2}/{R^2}$, where $B\equiv Eh^3/[12(1-\nu^2)]$ is the bending modulus~\cite{audoly2010elasticity}.
Both decrease with $R$, and their ratio ${\mathcal{E}_s}/{\mathcal{E}_b} \sim \left({d}/{h} \right)^2$ indicates that bending dominates for $d\ll h$, while stretching dominates for $d\gg h$.
Gravity adds an energy cost $\mathcal{E}_g\sim \rho g h R^2 d$, which prefers a smaller $R$.
Minimizing the total energy $\mathcal{E}_{\rm tot}=\mathcal{E}_s + \mathcal{E}_b + \mathcal{E}_g$ with respect to $R$ yields
\begin{align}
    \frac{R}{\ell_g} \approx 
    \begin{cases}
    c_b(\nu) \left( \frac{d}{h} \right)^{1/4} & (d\ll h) \\
    c_s(\nu) \left( \frac{d}{h} \right)^{3/4} & (d\gg h)
    \end{cases}, \label{sec03:eq:R}
\end{align}
where
\begin{align}
    \ell_g \equiv \left( \frac{B}{\rho g} \right)^{1/4} \label{sec03:def:ellg}
\end{align}
defines an ``{\it elasto-gravitational length},'' and $c_b$ and $c_s$ are numerical prefactors determined later.
Although the term ``elasto-gravitational length'' is often referred to as the characteristic wrinkling wavelength, $\lambda\sim (B/\rho_l g)^{1/4}$, of an elastic sheet floating on a liquid of density $\rho_l$~\cite{timoshenko1959theoryofplatesandshells, dillard2018review, box2017indentation}, the density $\rho$ considered in this study corresponds to that of the sheet itself.
The former length arises from a balance between the bending forces and buoyancy, whereas the latter is determined by the competition among bending, gravitational, and stretching forces.

The difference between the well-known \textit{gravito-bending length}~\cite{wang1986critical, cerda2004elements} $\ell_{\rm gb}=(B/(\rho g h))^{1/3}$ and the $\ell_g$ introduced here is significant. The former is the characteristic length that appears when the bending elasticity balances gravity such as drapes or beams bent substantially by their own weights. In our situation, however, stretching elasticity prevents large-amplitude deformations (until global buckling occurs). Therefore, the dominant length scale in our problem is given by $\ell_g$, which includes stretching elasticity.

Using $R$ in Eqs.~(\ref{sec03:eq:R}), we find that the total energy of the system scales $\rho g \ell_g^2h^{1/2}d^{3/2}$ when $d\ll h$, and $\rho g \ell_g^2h^{-1/2}d^{5/2}$ when $d\gg h$.
They must balance the work done during indentation, $Fd$.
Therefore we obtain
\begin{align}
    \frac{F}{\rho g h \ell_g^2} \approx 
    \begin{cases}
    k_b(\nu) \left( \frac{d}{h} \right)^{1/2} & (d\ll h) \\
    k_s(\nu) \left( \frac{d}{h} \right)^{3/2} & (d\gg h)
    \end{cases},  \label{sec03:eq:F}
\end{align}
where $k_b$ and $k_s$ are additional dimensionless prefactors.

Note that the scaling discussion here implicitly assumes that the sheet can be treated as a thin structure and that the indenter size is sufficiently small. A detailed examination of the case where $d$ is extremely small compared to $h$ reveals that the effect of the indenter shape becomes more pronounced, and the thin-plate approximation may no longer be valid. In the present study, we therefore focus on the regime in which the sheet can reasonably be regarded as thin ($d/h\gtrsim 10^{-1}$). To experimentally and numerically verify Eqs. ~(\ref{sec03:eq:R}) and (\ref{sec03:eq:F}), we restrict attention to the region where the indenter size is smaller than the representative lengths $\ell_g (d/h)^{1/4}$ and $\ell_g (d/h)^{3/4}$. For the elastomer used in our experiments ($\ell_g \sim 1$ mm), to observe the behavior in Fig.~\ref{Figure03} for $10^{-1} < d/h \lesssim 1$, it is desirable to use an indenter smaller than $\ell_g (d/h)^{1/4} \sim 1$ mm. Accordingly, we employed an indenter with radius $r_{\rm ind} = 0.69$~mm in the experiments. In the FES, we used $r_{\rm ind} = 0.1$~mm when $d/h < 10$, and a larger indenter of radius $r_{\rm ind} = 1.0$~mm for $d/h > 10$.  For $d\gg h$, the relation $R \sim \ell_g (d/h)^{3/4}$ ensures that $r_{\rm ind} / R \ll 1$, justifying the approximation of a point indenter~\cite{vella2017indentation}. Additional FES results showing the influence of $r_{\rm ind}$ are provided in the SM~\S~I.

The dashed and solid lines in Figure~\ref{Figure03} represent the predicted scaling laws~(\ref{sec03:eq:R}) and (\ref{sec03:eq:F}). Up to the global buckling threshold $d=d_c$, a wide range of data for elastic modulus, thickness, and even friction coefficient appears consistent with the predicted lines.

\section{Theoretical analysis of the axisymmetric state} \label{Sec:theory_liftedregion}

Our aim here is to analytically determine the pre-factors in the scaling relations Eqs.~(\ref{sec03:eq:R}) and (\ref{sec03:eq:F}).
We define a cylindrical polar coordinate $(r, \theta, z)$ and the origin is set to coincide with the center of the bottom surface of the sheet.
We introduce the displacement vector of the material points on the middle surface of the sheet as $\bm{u}(r, \theta) = u_r(r, \theta)\bm{e}_r + u_\theta(r, \theta)\bm{e}_\theta + w(r, \theta) \bm{e}_z$.
$w(r, \theta)$ represents the out-of-plane displacement, and the middle surface of the deformed sheet is represented by $z(r, \theta) = h/2 + w(r, \theta)$.

The mechanical equilibrium of a lifted sheet is described by the F\"{o}ppl-von K\'{a}rm\'{a}n (FvK) equations~\cite{audoly2010elasticity}: 
\begin{align}
    & B\nabla^4 w - h \sigma_{\alpha\beta} K_{\alpha\beta} + \rho g h = 0, \label{sec04:eq:FvKout}\\
    & h \nabla \cdot \sigma = 0, \label{sec04:eq:FvKin}
\end{align}
where $\sigma_{\alpha\beta}$ is components of the in-plane stress tensor, and the Greek indices run over $(r, \theta)$.
The stress and strain are related by Hookean linear constitutive equations, whereas the strain includes geometric nonlinearity in $w$ as $\epsilon_{rr}= u_{r,r} + w_{,r}^2/2$, 
$\epsilon_{r\theta}=\epsilon_{\theta r}= u_{r, \theta}/(2r) + u_{\theta, r}/2 - u_{\theta}/(2r) +  w_{,r}w_{,\theta}/(2r)$, and 
$\epsilon_{\theta\theta}= u_{r}/r + u_{\theta,\theta}/r +  w_{,\theta}^2/(2r^2)$, where $f_{,r}\equiv \partial f/\partial r$, and $f_{,\theta}\equiv \partial f/\partial \theta$~\cite{audoly2010elasticity}. 
$K_{\alpha\beta}$ is the curvature tensor, given by $K_{rr}=w_{,rr}$, $K_{r\theta}=K_{\theta r}=\partial_r(w_{,\theta}/r)$, and $~K_{\theta\theta}=w_{,r}/r + w_{,\theta\theta}/r^2$.
Equations~(\ref{sec04:eq:FvKout}) and (\ref{sec04:eq:FvKin}) represent the vertical and lateral force balances, respectively.

Assuming axisymmetry, the FvK equations can be reduced to the following ODEs:
\begin{align}
	& B\left[\frac{1}{r}\frac{d}{dr}\left(r\frac{d}{dr}\right)\right]^2 w - \frac{1}{r}\frac{d}{dr}\left( \psi \frac{dw}{dr}	\right) + \rho gh = 0, \label{sec04:eq:axiFvKout}\\
    & r\frac{d}{dr}\left[ 
    \frac{1}{r}\frac{d}{dr}(r\psi)\right]+\frac{Eh}{2}\left(\frac{dw}{dr}\right)^2 = 0, \label{sec04:eq:axicompat}
\end{align}  
where $\psi(r)$ denotes so-called the derivatives of the Airy stress function defined as $h\sigma_{rr}(r)=\psi/r,~h\sigma_{\theta\theta}(r)=d\psi/dr$~\cite{vella2018regimes}.
The in-plane equilibrium~(\ref{sec04:eq:FvKin}) is automatically satisfied by this representation, whereas $\psi(r)$ must obey the compatibility relation (\ref{sec04:eq:axicompat}).
The reaction force exerted at the center of a sheet equals to $F$, so that
\begin{align}
    F = 2\pi \left.\left\{ -\psi \frac{dw}{dr} + Br \frac{d}{dr}\left[\frac{1}{r}\frac{d}{dr}\left(r\frac{dw}{dr}\right)\right]\right\}\right|_{r=0}.
\end{align}
Using this relationship, we integrate Eq.~(\ref{sec04:eq:axiFvKout}), we obtain that
\begin{align}
	Br \frac{d}{dr}\left[\frac{1}{r}\frac{d}{dr}\left(r\frac{dw}{dr}\right)\right] - \psi \frac{dw}{dr} + \frac{1}{2} \rho gh r^2 = \frac{F}{2\pi}. \label{sec04:eq:axiFvKout_int}
\end{align}
Next, we analytically solve the coupled equations~(\ref{sec04:eq:axicompat}) and (\ref{sec04:eq:axiFvKout_int}).

\subsection{Small displacement:$d\ll h$}
We introduce the following dimensionless variables, motivated by the scaling relations Eqs.~(\ref{sec03:eq:R}) and (\ref{sec03:eq:F}) for $d\ll h$:
\begin{align}
    \xi \equiv \frac{r}{R},~~ W(\xi) \equiv\frac{w}{d}, ~~ \Psi(\xi) \equiv \frac{\psi R}{Ehd^2} ,~~ \mathcal{F} \equiv \frac{FR^2}{Bd}.  \label{sec04A:def:nondim}
\end{align}
With this nondimensionalization and assuming $R=c_b \ell_g (d/h)^{1/4}$, Eqs.~(\ref{sec04:eq:axiFvKout_int}) and (\ref{sec04:eq:axicompat}) become
\begin{align}
    &	\xi \frac{d}{d\xi}(\nabla_\xi^2 W) - 12(1-\nu^2) \left(\frac{d}{h}\right)^2 \Psi W' + \frac{c_b^4}{2} \xi^2 - \frac{\mathcal{F}}{2\pi} = 0, \label{sec04A:eq:FvKout}\\
    & \xi \frac{d}{d\xi} \left[\frac{1}{\xi}\frac{d}{d\xi}(\xi\Psi)\right] + \frac{1}{2 } W'^2 = 0, \label{sec04A:eq:compat}
\end{align}
where $f'\equiv df/d\xi$ and $\nabla_\xi^2 \equiv \frac{1}{\xi}\frac{d}{d\xi}\left(\xi\frac{d}{d\xi}\right)$.
Focusing on $d\ll h$, the second term in (\ref{sec04A:eq:FvKout}) can be neglected, and the equations for $W$ and $\Psi$, i.e., Eqs.~(\ref{sec04A:eq:FvKout}) and (\ref{sec04A:eq:compat}) are decoupled. 
The vertical displacement can then be determined from a single ODE as follows:
\begin{align}
    \xi \frac{d}{d\xi}(\nabla_\xi^2 W) + \frac{c_b^4}{2} \xi^2 - \frac{\mathcal{F}}{2\pi} = 0, \label{sec04A:def:FvKout_KL}
\end{align}
which can be solved exactly \cite{timoshenko1959theoryofplatesandshells, benson1991plate}.
We solve this by the following boundary conditions:
\begin{align}
    W(0)=1, ~ W'(0)=W(1)=W'(1)=W''(1)=0. \label{sec04A:sol:W}
\end{align}
The final condition is the moment-free condition at the detachment points $r=R$~\cite{vella2009statics, audoly2010elasticity, grandgeorge2022elastic}.
Solving this, we obtain
\begin{align}
    & W(\xi) = 1 - \xi^4 + 4\xi^2\ln{\xi}, \label{sec04A:sol:cb} \\
    & c_b = 2^{3/2}, \label{sec04A:sol:kb}\\
    & k_b =4{\pi}.
    \label{sec03:eq:Fd_bend_w_pre}
\end{align}
In Fig.~\ref{Figure03}, we compared Eqs.~(\ref{sec03:eq:R}), (\ref{sec03:eq:F}), (\ref{sec04A:sol:kb}), and (\ref{sec03:eq:Fd_bend_w_pre}) with the experimental and FES data. Despite the simplifying assumption of the point-load indentation, we observe that our analytical predictions show reasonable agreement with numerical and experimental data obtained for finite size indenters.

\subsection{Large displacement: $d\gg h$}

Next, we introduce the following dimensionless variables:
\begin{align}
    \xi \equiv \frac{r}{R},~~ W(\xi) \equiv\frac{w}{d}, ~~ \Psi(\xi) \equiv \frac{\psi R}{Ehd^2} ,~~ \mathcal{F} \equiv \frac{FR^2}{Ehd^3},
\label{sec04B:def:nondim}
\end{align}
where $R=c_s \ell_g (d/h)^{3/4}$.
For these variables, Eq.~(\ref{sec04:eq:axiFvKout}) becomes
\begin{align}
  \frac{1}{12(1-\nu^2)}\left(\frac{h}{d}\right)^2 \xi \frac{d}{d\xi}(\nabla_\xi^2 W) - \Psi W' + \frac{c_{s}^4}{24(1-\nu^2)}\xi^2  - \frac{\mathcal{F}}{2\pi} = 0, 
\end{align}
and considering $d \gg h$, Eqs.~(\ref{sec04:eq:axiFvKout_int}) and (\ref{sec04:eq:axicompat}) are reduced to
\begin{align}
  &  - \Psi W' + \frac{\alpha}{2} \xi^2 - \frac{\mathcal{F}}{2\pi} = 0,  \label{sec04B:eq:Memb} \\
  &      \xi \frac{d}{d\xi} \left[\frac{1}{\xi}\frac{d}{d\xi}(\xi\Psi)\right] + \frac{1}{2 } W'^2 = 0, \label{sec04B:eq:compat}
\end{align}
where $\alpha \equiv c_s^4/(12(1-\nu^2))$.

Exact analytical solutions of Eqs.~(\ref{sec04B:eq:Memb}) and (\ref{sec04B:eq:compat}) are known in $\alpha=0$, and take a particularly simple closed form for the special value of the Poisson's ratio $\nu=1/3$~\cite{mansfield1989bending, chopin2008liquid}.
However, for $\alpha > 0$, they do not seem to admit any concise analytical solutions.
As we can estimate $\alpha\sim 0.3$ from our experimental and FES results, we treat the second term in the left-hand side of Eq.~(\ref{sec04B:eq:Memb}) as a perturbative term.
Therefore, we use the closed-form solution at $\alpha=0$ as the unperturbed state and compute $O(\alpha)$ corrections by expanding $W$, $\Psi$, and $\mathcal{F}$ in $\alpha$, subject to the boundary conditions
\begin{align}
& W(0)=1,~W(1)=0,~\Psi(0)=0,~\Psi'(1)-\nu\Psi(1)=0, \label{sec04B:bc}
\end{align}
with $\nu=1/3$.
The last condition corresponds to a clamped radial displacement at $r=R$, i.e., $u_r(R)=0$.
Although this condition is not satisfied exactly in our system (see below), it is still reasonable for our present purpose, because the radial displacement at the detachment radius remains small compared to other length scales (for example, Fig.~\ref{Figure05} shows $|u_r(R)|< h$).

The perturbation calculation yields the approximate solutions for $W$, $\Psi$, and $\mathcal{F}$ as a series in $\alpha$.
We then determine the unknown constant $\alpha$ by minimizing $\mathcal{E}_s + \mathcal{E}_g$ with respect to $R$, which yields
\begin{align}
    & c_s \approx \left(\frac{32}{9}\right)^{1/4} \approx 1.373, \label{sec04B:sol:cs} \\
    & k_s \approx \frac{5\sqrt{2}\pi}{3}\approx 7.405.\label{sec04B:sol:ks}
\end{align}
The full derivation is provided in SM~\S~II.
Equations~(\ref{sec03:eq:R}) and (\ref{sec03:eq:F}) with the prefactors (\ref{sec04B:sol:cs}) and (\ref{sec04B:sol:ks}) are plotted as solid lines in Fig.~\ref{Figure03}, and show excellent agreement with both the experiment and FES.

\begin{figure}[]
    \centering	\includegraphics[width=1.\linewidth]{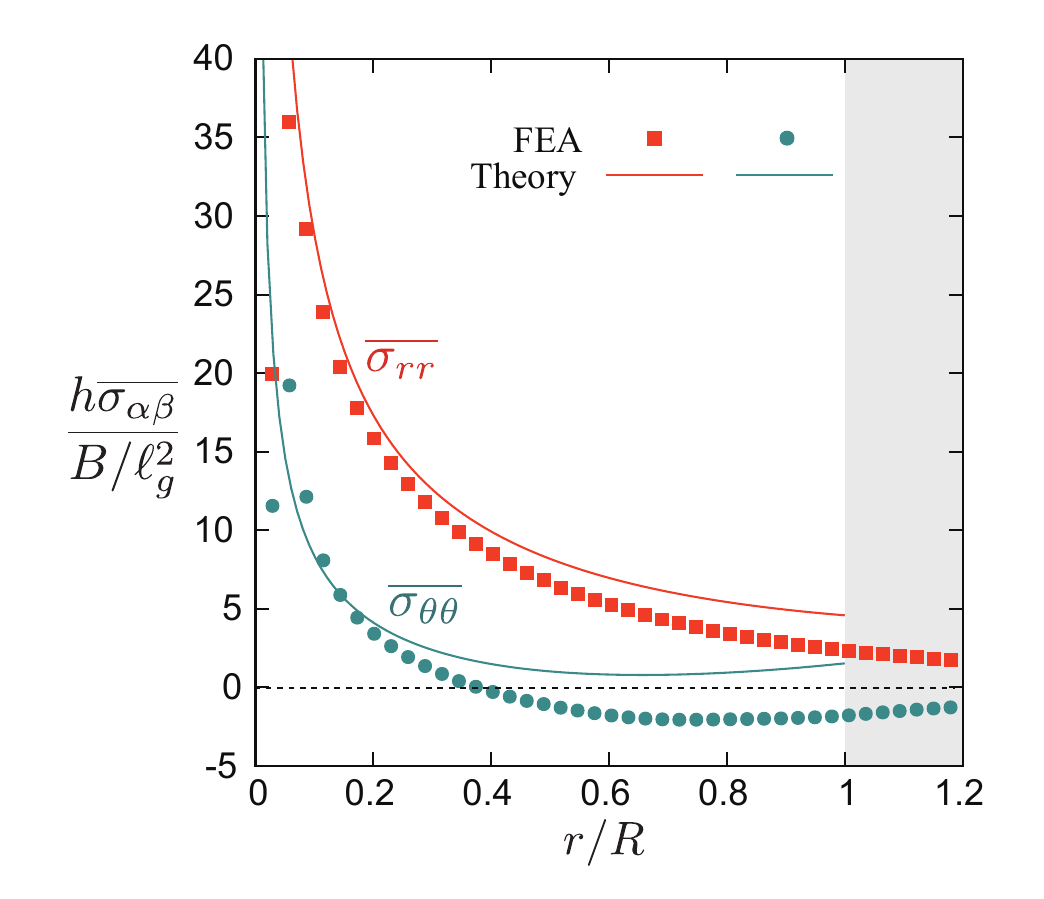} 	\caption{
    The profile of the azimuthally averaged stress $h\overline{\sigma_{\alpha\beta}}(r)=\frac{1}{2\pi}\oint h\sigma_{\alpha\beta}(r, \theta) d\theta$ is plotted normalized by the characteristic stress scale $B/\ell_g^2$.
    Dots and solid lines represent a FES result based on the simulation described in Fig.~\ref{Figure01} (with $d=4.66~\textrm{mm}$) and our theory in \S~\ref{Sec:theory_liftedregion}, respectively.
    The shaded region corresponds to the contact region. 
 }\label{Figure04}
\end{figure}

From this analysis, we also obtained the in-plane stresses $\sigma_{rr}$ and $\sigma_{\theta\theta}$ in the lifted region (see SM~\S~II).
However, the discrepancy between these analytical stress profiles and the FES becomes significant near $r=R$ [see Fig.~\ref{Figure04}].
Although the analysis based on the clamped boundary condition, $u_r(R)=0$, provides good estimations of $F$ and $R$, it fails to predict a detailed stress distribution~\cite{chopin2008liquid, dai2021poking}.
Because the center of the sheet is pushed upwards, the entire region of $0<r<R$ must be stretched both radially and azimuthally under the clamped boundary condition. Consequently, the present theoretical analysis predicts the positive stress field over the entire considered region. In contrast, both in the indentation test and FES, the material is allowed to slide on the substrate at $r=R$. As $d$ increases, material points undergo a small inward radial displacement, and the azimuthal direction contracts, leading to the appearance of the circumferentially compressive region $\sigma_{\theta\theta}<0$.
Thus a more accurate analytical description of the stress field requires matching the solution of the FvK equations for $r<R$ with a planar solution for $r>R$~\cite{chopin2008liquid, vella2018regimes, dai2020radial}. 
We leave this matching problem for future work, and we here will focus on the fundamental properties of the stress field within the contact region.

%%% FIGURE 4

\section{Displacement and stress in the contact region}

In previous studies on the indentation-induced wrinkling of thin sheets~\cite{chopin2008liquid, box2017indentation, vella2018regimes}, the emergence of wrinkles has been attributed to the destabilization of axisymmetric configurations once the compressive azimuthal in-plane stress exceeds a buckling threshold.
Here we are particularly interested in the role of contact forces in wrinkle formation.
Therefore, we focus on the in-plane displacement field and the stress distribution within the contact region, and we reveal how friction affects the hoop stress distribution.

In the following, we often use azimuthally averaged quantities defined as $\overline{f}(r) \equiv (2\pi)^{-1}\oint f(r, \theta)\, d\theta$, for example, when characterizing the displacement and stress fields. Numerically, $\overline{f}(r')$ is computed by discretizing $r$ and averaging $f$ over all nodes within a small radial interval $r' - \Delta r < r < r' + \Delta r$. 
When the system exhibits a high degree of axisymmetry, $\overline{f}(r)$ provides a useful representation of the spatial variation. However, it should be noted that for globally buckled structures, in which axisymmetry is broken, $\overline{f}(r)$ should be regarded only as the quantity for reference.

\subsection{Displacement}

\begin{figure*}[]
    \centering	\includegraphics[width=1.\linewidth]{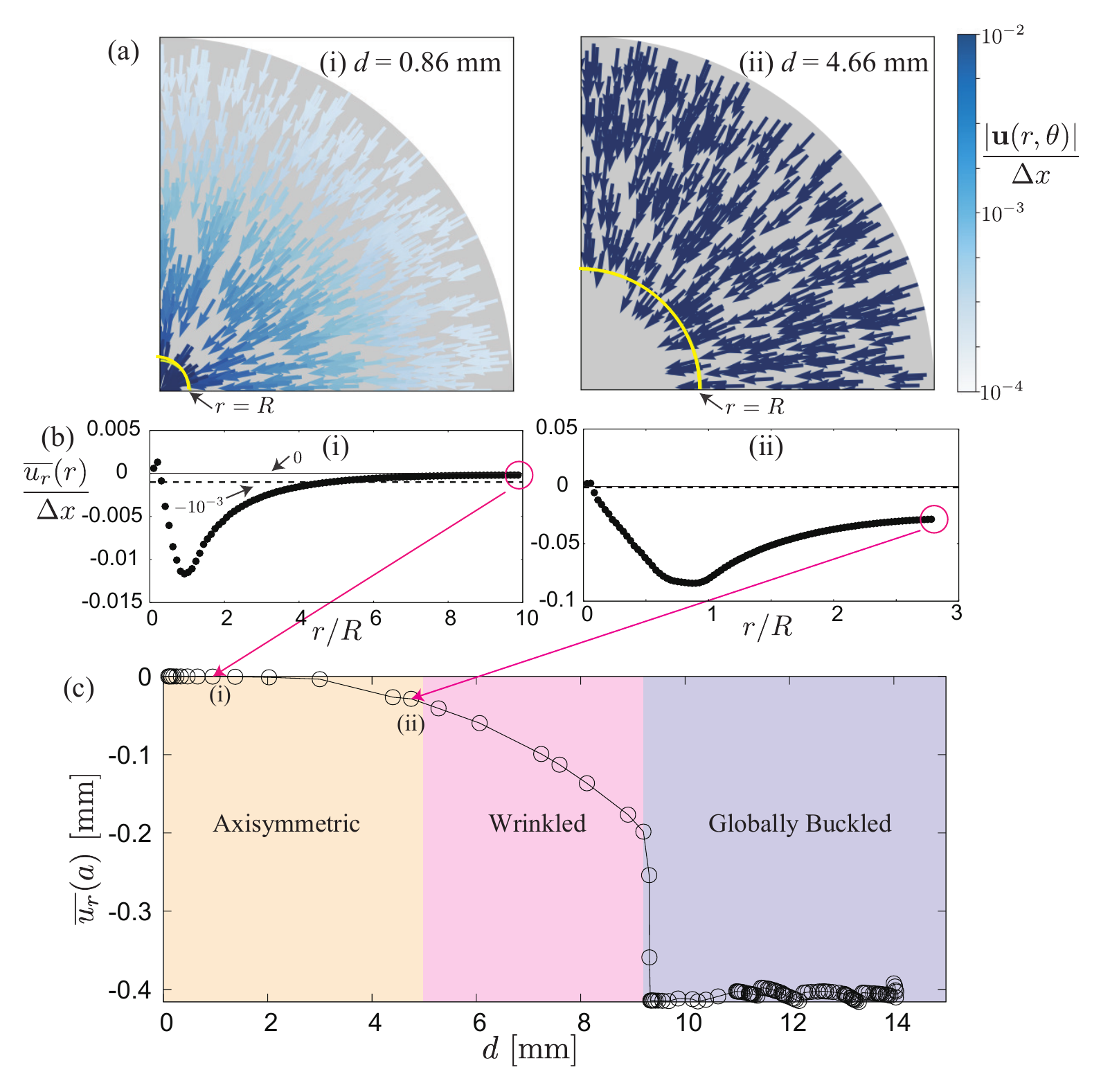} 	\caption{
    Displacement field in the contact region for (i) small indentation height ($d = 0.86$~mm) and (ii) large indentation height ($d = 4.66$~mm). 
    (a) Quarter of the sheet is shown as the shaded area. The color of the inward-pointing vectors represents the magnitude of the displacement $\bm{u}(r, \theta)$ in the contact region, normalized by the representative mesh size $\Delta x$~(here equals to $1~\textrm{mm}$) of the finite elements. 
    (b) Radial profiles of the azimuthally averaged radial displacement, defined as $\overline{u_r}(r) \equiv \frac{1}{2\pi} \oint u_r(r, \theta) d\theta$, plotted as a function of $r/R$. 
    The solid and dashed reference lines indicate $\overline{u_r}/\Delta x = 0$ and $-10^{-3}$, respectively. 
    (c) Radial displacement at the edge of the sheet, $\overline{u_r}(a)$, as a function of $d$. 
    The values of $\overline{u_r}(a)$ are extracted from the red circle markers in panel (b). 
    $\overline{u_r}(a)/\Delta x\gtrsim 10^{-3}$ indicates that the entire sheet has slipped. 
    In the parameter ranges explored in this study, almost all of the sheet slip completely before the onset of wrinkling. 
    These results are obtained from the finite-element simulations shown in Fig.~\ref{Figure01}.
     }\label{Figure05}
\end{figure*}

We first examine the in-plane displacement using the simulation data shown in Fig.~\ref{Figure01} (with a finite value of $\mu=0.32$).  
The displacement vector $\bm{u}(r, \theta)$ in the contact region is shown in Fig.~\ref{Figure05}(a).  
The color of the arrows represents $|\bm{u}(r, \theta)| / \Delta x$, where $\Delta x$ is the representative mesh size.
We observe inward slip within the contact region as the indentation proceeded, most prominently near $r\sim R$.

For sufficiently small $d$, a region of moderately large displacements ($|\bm{u}|/\Delta x \gtrsim 10^{-3}$) exists, whereas the displacement in the surrounding area is much smaller than the mesh size $\Delta x$ [Fig.~\ref{Figure05}(a--i)]. The fact that $|\bm{u}|$ is extremely small compared to $\Delta x$ may suggest that the shear stress between the substrate and the sheet does not exceed the maximum static frictional force (per unit area).
Based on this observation, we assume the threshold for determining whether or not the sheet slides with respect to the substrate as $|\bm{u}|/\Delta x = 10^{-3}$.
Fig.~\ref{Figure05}(b-i) shows the azimuthally averaged radial displacement $\overline{u_r}(r)$, which reveals that the interface between regions of ``large'' and ``small'' displacement lies near $r/R \sim 5$.
As $d$ increases, this interface moves outward, and eventually the entire contact region undergoes a substantial inward slip [see Fig.~\ref{Figure05}(a-ii) and (b-ii); see also SM~\S III].  
We refer to this state as {\it complete slipping}.

We also plot the azimuthally averaged radial displacement at the sheet edge, $r = a$, in Fig.~\ref{Figure05}(c).  
This indicates that the contact region has already reached a completely slipping state at the onset of wrinkling.
This scenario, in which wrinkle formation occurs after the onset of complete slipping, is observed for a wide range of parameter values considered in this study.
The critical displacement at the onset of complete slipping is further discussed in SM~\S III.
Accordingly, in the following analysis, we restrict our theoretical investigation to the stress field at the contact surface in the case of complete slipping.

\subsection{Stress}

The indentation speed in our experiments is quite low ($\dot{d}=0.1$~mm/s); therefore we proceed the theoretical analysis with the quasi-static limit $\dot{d}\rightarrow 0$.
In the Abaqus simulations, we use the default Coulomb friction model, in which the interfacial shear traction saturates at the Coulomb threshold $\mu f_n$ and is treated as independent of sliding velocity once slip has initiated~\cite{abaqusdoc}, where $f_n$ represent the normal force.
Although in principle, rate- and state-dependent friction can lead to a small difference between the friction coefficient at the onset of sliding and that during slow steady sliding~\cite{baumberger2006solid,popov2010contact}, these effects are expected to be negligible under the present loading protocol.
It is therefore reasonable, both in our experiments and in the corresponding Abaqus modeling, to treat the interfacial tangential traction as being everywhere close to the Coulomb threshold, with an effective magnitude $f_r \approx \mu \rho g h$, which we regard as spatially uniform over the contact surface.
Under this assumption, the axisymmetric in-plane force balance~\cite{timoshenko1969theoryofelasticity} within the contact region is 
\begin{align}
&    \frac{\sigma_{rr} -\sigma_{\theta\theta}}{r} + \frac{d}{dr}\sigma_{rr} + \mu \rho g = 0. ~\label{sec05B:eq:Lame}
\end{align}
This equation can be solved once the boundary values $\sigma_{rr}(a)=0$ and $\sigma_{rr}(R)$ are specified.
In the frictionless case ($\mu=0$), we immediately obtain a solution of $\sigma_{\theta\theta}(r)\sim -\sigma_{rr}(R) r^2/R^2$.
Such a negative stress field cannot be matched smoothly at $r=R$ with the clamped solution for $r<R$ derived from \S~\ref{Sec:theory_liftedregion}.
Instead, Fig.~\ref{Figure04} shows that the FES results exhibit $\sigma_{\theta\theta}<0$ near $r\approx R$, consistent with the above prediction.
The negative hoop stress at $r=R$ arises because the vertical deflection pulls material radially inwards ($u_r(R)<0$), and it can destabilize the axisymmetric state~\cite{chopin2008liquid, box2017indentation, vella2018regimes, dai2018interface}.

\begin{figure}[]
    \centering	\includegraphics[width=.9\linewidth]{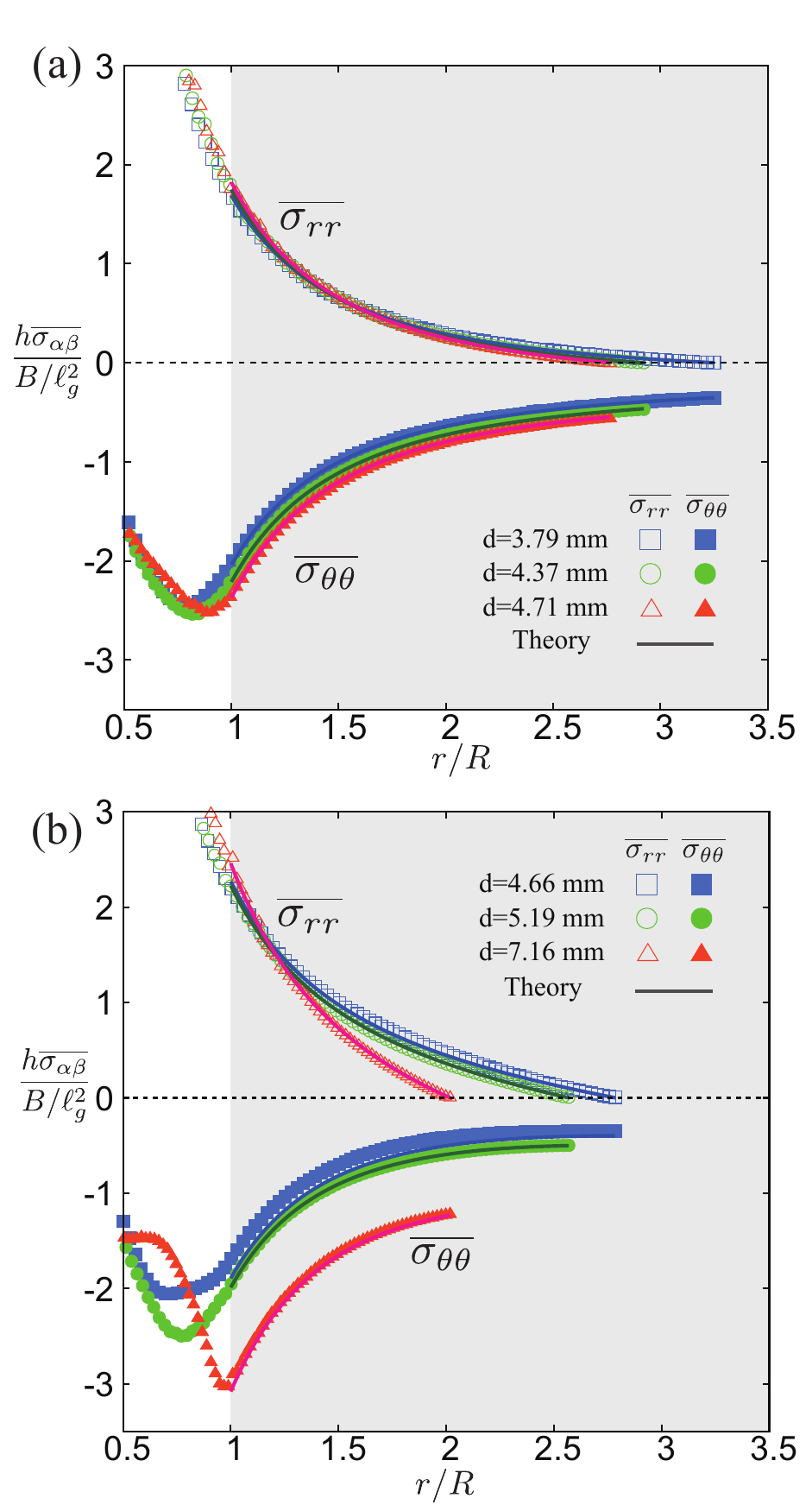} 	\caption{
    Azimuthally averaged in-plane stress profiles obtained from FES.
    Open and closed symbols represent the radial and hoop components of the azimuthally averaged stress tensor, respectively.
    The solid line indicates the planar solution of the force balance equation~(\ref{sec05B:eq:Lame}) with $\sigma_{rr}(R)$ determined from fitting to the FES data.
    The shaded region corresponds to the contact region.
    (a) The results without friction $\mu=0$, with all the other parameters identical to those in Fig.~\ref{Figure01}.
    (b) The results obtained from the same parameter set as that shown in Fig.~\ref{Figure01} with $\mu=0.32$.
    The critical displacements $d_w$ in (a) and (b) are 4.4 and 4.9 mm, respectively.
 }\label{Figure06}
\end{figure}

The analytical solutions of Eq.~(\ref{sec05B:eq:Lame}) are provided in Eq.~(S45) and (S46) in the Supplementary material; here, we present a simpler approximation valid for $r\sim R$ and $R\ll a$:
\begin{align}
  &  \sigma_{rr}(r)\sim \sigma_{rr}(R)\frac{R^2}{r^2}, \label{sec05B:eq:srr} \\
  &  \sigma_{\theta\theta}(r)\sim -\sigma_{rr}(R)\frac{R^2}{r^2} + \mu \rho g a.
  \label{sec05B:eq:stt}
\end{align}
We fit the FES data using the exact expressions~(S45) and (S46), treating $\sigma_{rr}(R)$ as a fitting parameter.
The fitted value of $\sigma_{rr}(R)$ was found to be approximately half of that predicted from the clamped solution for $r<R$ in \S~\ref{Sec:theory_liftedregion}.
The results are presented in Fig.~\ref{Figure06}. 
Our solution reproduced the FES stress profiles well for $r>R$, both in the frictionless case ($\mu=0$) and with friction ($\mu=0.32$), even after wrinkle formation.

Equations~(\ref{sec05B:eq:srr}) and (\ref{sec05B:eq:stt}) show that, near $r\sim R$, the radial stress $\sigma_{rr}$ is only weakly affected by friction, whereas the magnitude of the compressive hoop stress $\sigma_{\theta\theta}$ is reduced by the $\mu\rho g a$ term.
The frictional dependence of $\sigma_{rr}(R)$ is shown in SM \S~IV.

%%% FIGURE 5

%%% FIGURE 6

\section{Onset of wrinkling instability} \label{sec:wrinkling}

Building on the previous sections, in which we clarified the stress fields in both the lifted and contact regions, we now investigate how these in-plane stresses contribute to wrinkle formation through a linear stability analysis of the FvK equations~\cite{davidovitch2011prototypical, box2017indentation}.

We consider a perturbative solution of the FvK equations describing a sinusoidal out-of-plane variation with $m$ wrinkles and a small amplitude $f(r)$, added to the axisymmetric base state $w^{(0)}(r)$:
\begin{align}
    w(r, \theta) = w^{(0)}(r) + f(r) \cos(m\theta).\label{sec06:eq:w_wrinkle}
\end{align}
Substituting Eq.~(\ref{sec06:eq:w_wrinkle}) into Eq.~(\ref{sec04:eq:FvKout}), and assuming axisymmetric stress fields, the equation for $f$ is given by
\begin{align}
        B\left[ \frac{1}{r}\frac{d}{dr} \left( r\frac{d}{dr} \right) - \frac{m^2}{r^2} \right]^2 f(r) - h \sigma_{rr} \frac{d^2f(r)}{dr^2} \nonumber \\
        - h \sigma_{\theta\theta} \left( -\frac{m^2}{r^2} + \frac{1}{r}\frac{d}{dr} \right) f(r) = 0. \label{sec06:eq:FvK_wrinkle}
\end{align}
Fig.~\ref{Figure06} indicates that $|\sigma_{\theta\theta}|$ reaches its maximum near $r=R$. 
Given that this large compressive stress is essential for wrinkle formation, we estimate the magnitude of each term in Eq.~(\ref{sec06:eq:FvK_wrinkle}) in the vicinity of $r=R$~\cite{davidovitch2011prototypical}.
The first term in the l.h.s of Eq.~(\ref{sec06:eq:FvK_wrinkle}) comes from the bending elasticity $\sim Bm^4f(R)/{R^4}$, whereas the third term $\sim h |\sigma_{\theta\theta}(R)| m^2 f(R)/{R^2}$ represents the azimuthal compression.
The second term, which is the product of the radial tension and curvature, $\sim h\sigma_{rr}(R)f(R)/R^2$ acts as a Laplace-pressure-like restoring force when $|\partial_r^2f|>0$.
This is analogous to that of a compressed elastic beam resting on an elastic foundation~\cite{timoshenko1959theoryofplatesandshells, dillard2018review} with an effective stiffness $K_{\rm eff}\sim h \sigma_{rr}(R)/R^2$.
By balancing the restoring force from the foundation ($\sim K_{\rm eff}f(R)$) with the bending term, we obtain the optimal wavelength as $\lambda\sim (B/K_{\rm eff})^{1/4}$~\cite{cerda2003geometry}.

%%% TABLE 2
\begin{table}
% \begin{table*}
 \begin{center}
   \caption{
   Properties of the elastic sheets used to investigate the onset of wrinkling instability.
    The table summarizes the method of investigation (experiment or FEM), material type, sheet radius $a$, sheet thickness $h$, and the coefficient of static friction $\mu$.
    It also indicates the data markers used to represent each elastic sheet in Figs.~\ref{Figure07}, \ref{Figure08} and \ref{Figure09} for both experimental and numerical results.
   }
   \label{Table02}
\includegraphics[width=1.\linewidth]{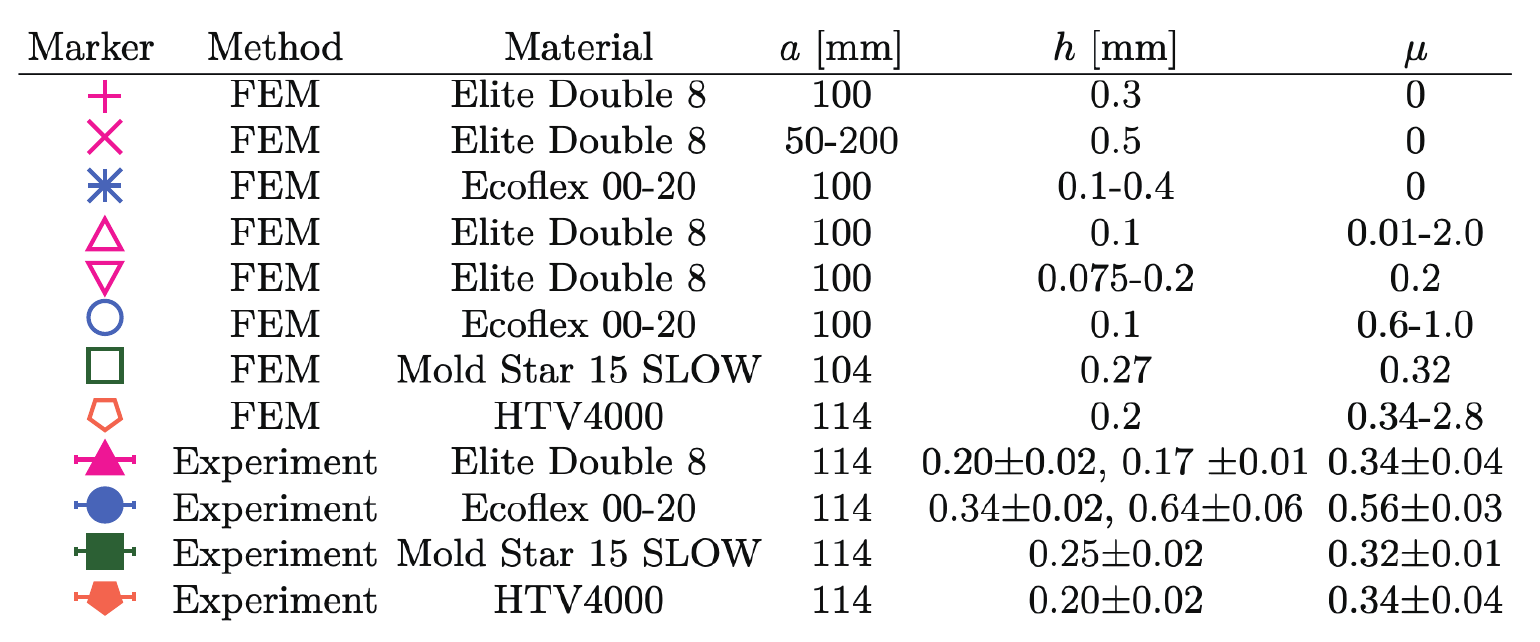} 
 \end{center}
% \end{table*}
\end{table}

\subsection{Frictionless case ($\mu=0$)}

In the frictionless case ($\mu=0$), according to Eq.~(\ref{sec03:eq:R}), $\sigma_{rr}\sim |\sigma_{\theta\theta}|\sim E(d/R)^2\sim B/(h\ell_g^2)(d/h)^{1/2}$.
The balance between the restoring and azimuthal compressive force in Eq.~(\ref{sec06:eq:FvK_wrinkle}) yields $m \sim (\sigma_{rr}(R)/|\sigma_{\theta\theta}(R)|)^{1/2}\sim O(1)$.
Using $m =2\pi R/\lambda \sim (K_{\rm eff}/B)^{1/4}R$, as derived in the previous paragraph, together with the scaling relations for $\sigma_{rr}(R)$ and $R$, we obtain $m \sim (d/h)^{1/2}$.  
With $m\sim O(1)$, this gives $d/h\sim m^2\sim O(1)$, i.e., the wrinkling threshold depends only on the sheet thickness, $d_w\propto h$.

\begin{figure}[]
    \centering	\includegraphics[width=1.\linewidth]{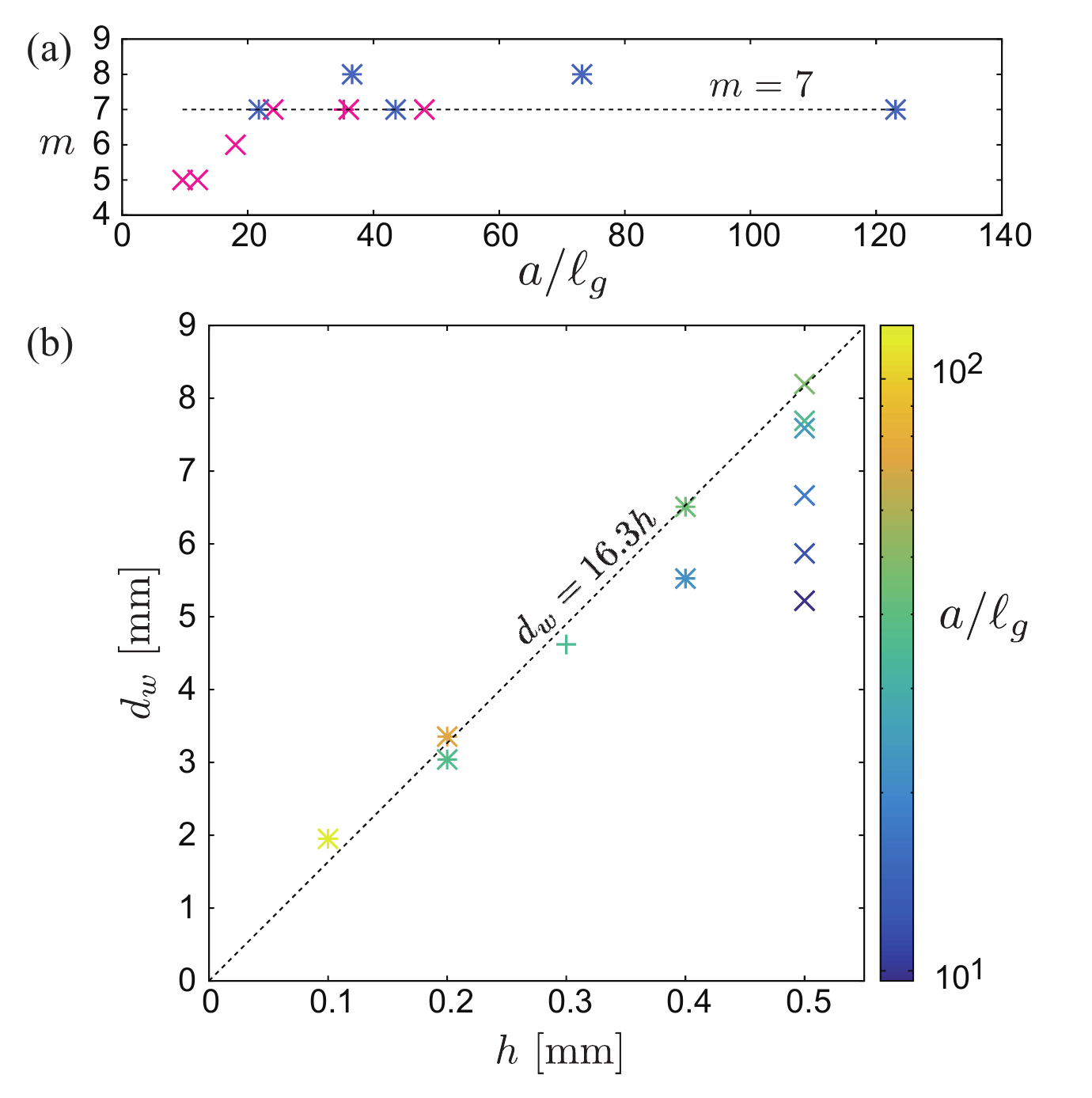} 	\caption{
    The number of wrinkles and the critical displacement at the onset of wrinkling instability in the frictionless case.
    Details of each data point are summarized in Table~\ref{Table02}.
    (a) Number of wrinkles $m$ as a function of $a/\ell_g$.
    For $a/\ell_g \gtrsim 20$, all data points are located near $m=7$.
    (b) Relationship between the critical displacement $d_w$ and sheet thickness $h$.
    Color indicates the value of $a/\ell_g$.
    The dashed lines correspond to the expression given in Eqs.~(\ref{sec06A:eq:m0}) and (\ref{sec06A:eq:dw_0}).
 }\label{Figure07}
\end{figure}

To verify the above prediction, we performed FES for the case of $\mu = 0$ using the parameters listed in Table~\ref{Table02}.
We investigated the critical displacement at the onset of the wrinkling, $d_{w}$ and the number of wrinkles, $m$ using a fast Fourier transformation of the node profile, $w(R, \theta)$ (using \texttt{numpy.fft} in Python).
The results are presented in Fig.~\ref{Figure07}.
When the sheet radius $a$ is sufficiently larger than the elasto-gravitational length $\ell_g$, the number of wrinkles typically ranges from seven to eight, with $m=7$ being the most frequent, and $d_w$ is proportional to $h$ with a slope of 16.33.
By combining our theoretical analysis and FES results, we conclude that
\begin{align}
    m^0 &\approx 7, \label{sec06A:eq:m0} \\
    \frac{d_{w}^0}{h} &\approx 16.33, \label{sec06A:eq:dw_0}
\end{align}
where the subscript ``0'' indicates that $\mu=0$.
The constants on the right-hand side are universal, meaning that they are independent of any material parameters, such as the Young's modulus of the sheet.

As $a/\ell_g$ decreased, the number of wrinkles tended to decrease to five and the deviation in Fig.~\ref{Figure07}~(b) became more pronounced.
This trend is likely due to the finite-size effects of the sheet, which were not taken into account in the theoretical analysis above.
We will return to this point in Sec.~\ref{sec:globalbuckling}.

%%% FIGURE 7

\subsection{Frictional case ($\mu>0$)}

\begin{figure}[]
    \centering	\includegraphics[width=1.\linewidth]{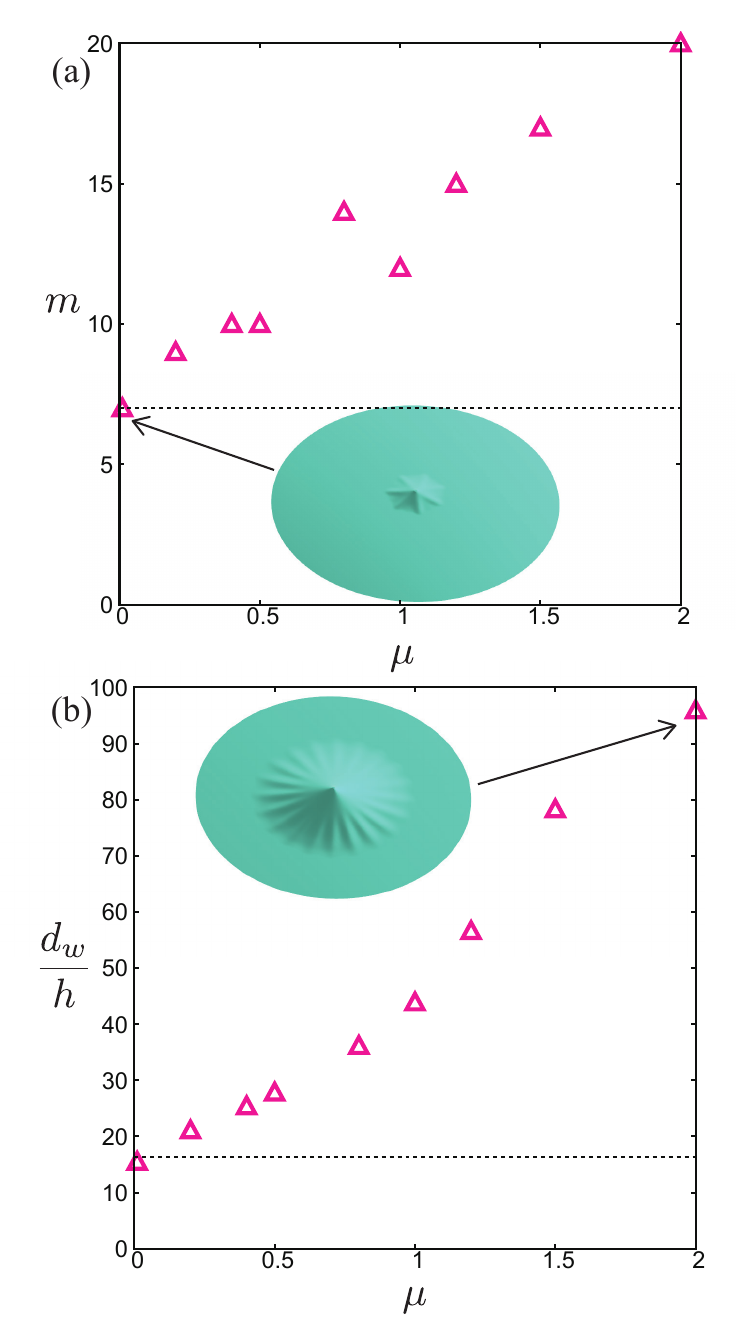}
    \caption{
    The number and critical displacement of wrinkles in the frictional cases, obtained from FES for varying friction coefficients $\mu$, keeping all other parameters fixed.
    Both the critical number of wrinkles $m$ [panel(a)] and the rescaled critical displacement $d_w/h$ [panel(b)] increase monotonically with $\mu$.
    The dotted lines in (a) and (b) indicate those without friction $\mu=0$, given in Eqs.~(\ref{sec06A:eq:m0}) and~(\ref{sec06A:eq:dw_0}).
    The inset images show the wrinkled shapes in FES for $\mu = 0.01$ and $\mu = 2.0$. The parameters used are those of ``Elite double 8" given in Table~\ref{Table02}.
 }\label{Figure08}
\end{figure}

When frictional interactions occur between the sheet and the substrate, $d_w/h$ and $m$ are no longer constant and they increase monotonically with $\mu$ as shown in Fig.~\ref{Figure08}. 
Qualitatively, this can be understood as follows: friction suppresses $|u_r(R)|$, which in turn reduces $|\epsilon_{\theta\theta}(R)|$ compared to the frictionless case at the same $d$; therefore, a larger $d$ is required to destabilize the axisymmetric state~\cite{dai2020radial, chawla2024geometry}. 

As shown in Eqs.~(\ref{sec05B:eq:srr}) and (\ref{sec05B:eq:stt}), the contribution of friction to the hoop stress can be estimated as $\sim \mu \rho g a$, whereas its effect on the radial stress appears only at a higher orders.
By normalizing the contraction force $h|\sigma_{\theta\theta}|$ with the characteristic stress scale $B/\ell_g^2(=\sqrt{B\rho g})$, we find the {\it relative frictional stress}:
\begin{align}
    \tau \equiv \frac{\mu a h}{\ell_g^2}, \label{sec06B:def:tau}
\end{align}
which is a dimensionless measure of the frictional contribution to the hoop stress.
As we demonstrate below, this parameter $\tau$ plays a central role in characterizing  the frictional effects in our problem.

\begin{figure*}[]
    \centering	\includegraphics[width=1.\linewidth]{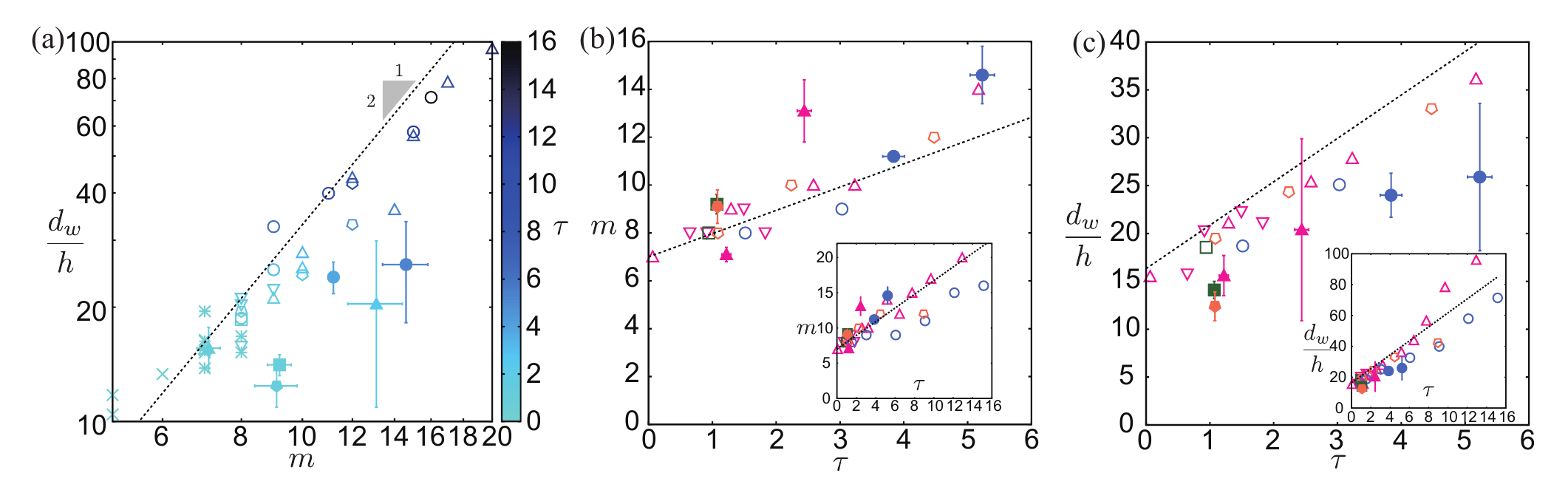} 	\caption{
    Relationships among the number of wrinkles $m$, the normalized critical displacement $d_w/h$, and the normalized frictional stress $\tau$ defined in Eq.~(\ref{sec06B:def:tau}).
    Details of the symbols are summarized in Table~\ref{Table02}.
    (a) Relationship between $d_w/h$ and $m$.
    The dashed line corresponds to the expression given in Eq.~(\ref{sec06B:eq:dw_m}).
    The color bar represents the magnitude of $\tau$.
    (b, c) The number of wrinkles $m$ and the normalized displacement $d_w/h$ for $\mu > 0$, plotted as functions of $\tau$.
    For small $\tau$, the rescaled data are captured by the master curves described by Eqs.~(\ref{sec06B:eq:m_fric}) and (\ref{sec06B:eq:dw_fric}), with the common fitting parameter $c = 0.14$.
    Data over a wide range of $\tau$ are also shown in the insets of these figures.
 }\label{Figure09}
\end{figure*}

We now refine the preceding scaling relations in Eqs.~(\ref{sec06A:eq:m0}) and (\ref{sec06A:eq:dw_0}), by including frictional effects up to the first order in $\tau$.
The scaling relation $d_w/h \sim m^2$, obtained by balancing the bending force and the radial tension, is expected to remain valid even for $\mu > 0$, because $\sigma_{rr}$ is insensitive to $\mu$.
Indeed, FES performed using the parameter sets listed in Table~\ref{Table02} confirms that the scaling prediction
\begin{align}
    \frac{d_w}{h} \approx 0.33 m^2 \label{sec06B:eq:dw_m}
\end{align} 
agrees well with the FES results [Fig.~\ref{Figure09}(a)].
Here, the numerical coefficient 0.33 is determined within the scaling argument for $\mu=0$; that is, $(d_w^0 /h)/ (m^0)^2 \approx 0.33$ based on Eqs.(\ref{sec06A:eq:m0}) and (\ref{sec06A:eq:dw_0}).

Based on the balance between radial and hoop stresses in Eq.~(\ref{sec06:eq:FvK_wrinkle}), we obtain $m \sim (\sigma_{rr}(R)/|\sigma_{\theta\theta}(R)|)^{1/2} \sim 1 + O(\tau)$, which is approximately equal to $(d_w/h)^{1/2}$ since Eq.~(\ref{sec06B:eq:dw_m}) holds.
Therefore, we obtain the following:
\begin{align}
& m \approx m^0 \left( 1 + c \tau \right), \label{sec06B:eq:m_fric}\\
& \frac{d_{w}}{h} \approx \frac{d_{w}^0}{h} \left( 1 + c \tau\right)^2 \approx \frac{d_{w}^0}{h} \left( 1 + 2c \tau\right), \label{sec06B:eq:dw_fric}
\end{align}
where $c$ is a numerical constant common to both the expressions.

To test Eqs.~(\ref{sec06B:eq:m_fric}) and (\ref{sec06B:eq:dw_fric}), we performed additional experiments.
In our experiments, visually detecting the presence or absence of wrinkles was challenging, and the force–displacement curve did not provide any information related to wrinkling.
Thus, we determined the presence of wrinkles by carefully observing the point-cloud data of the sheet obtained via 3D scanning.
From our measured data, we identified the maximum displacement for the axisymmetric state, $d_w^{-}$ and the minimum displacement for the wrinkled state,  $d_w^{+}$, respectively, and determined the critical displacement in our experiments as $d_w\equiv (d_w^{+}+d_w^{-})/2$.
The number of wrinkles $m$ was determined from wave profiles: for several radii $r$ we analyzed the angular dependence of $w(r,\theta)$ in the point-cloud data and computed a mean number of wrinkles from $m = 2\pi r / \lambda(r)$ for each sample.
In our experiments, the wave profiles were occasionally unevenly distributed along the circumferential direction.
The irregular wavelengths $\lambda$ may have originated from small defects, such as impurities introduced during sample fabrication, and/or from residual pre-stress induced when the sheet was placed on the substrate.
Indeed, we did not observe such a irregularities in the wave profiles in our FES.
We prepared four to five samples from the same material and with similar thicknesses (thickness variation within 10\%, as shown in Table~\ref{Table02}).
We performed the above analysis for each sample, and calculated the mean and standard deviation for each group with similar thickness.

Our theoretical predictions, experimental data, and FES results are summarized in Fig.~\ref{Figure09}.
The parameter sets investigated in the experiments and FES, together with the corresponding symbols, are listed in Table~\ref{Table02}.
The dashed lines represents the theoretical predictions of Eq.~(\ref{sec06B:eq:dw_m}), (\ref{sec06B:eq:m_fric}), and (\ref{sec06B:eq:dw_fric}), which consist of three dimensionless parameters $m^0$, $d_{w}^0/h$, and $c$.
We fixed $m^0$ and $d_{w}^0/h$ as the frictionless values given in Eqs.~(\ref{sec06A:eq:m0}) and (\ref{sec06A:eq:dw_0}), respectively.
The remaining parameter $c$ was determined by fitting the FEM data in Fig.~\ref{Figure09} (b) to Eq.~(\ref{sec06B:eq:m_fric}) for $\tau<2$, yielding $c\approx 0.14$.
We use this value of $c$ for the dashed curve in Fig.~\ref{Figure09} (c).
The scaling plots of our data for a wide range of values of $E$, $h$, and $\mu$ in Fig.~\ref{Figure09} supports our predictions, Eqs.~(\ref{sec06B:eq:m_fric}) and (\ref{sec06B:eq:dw_fric}), up to moderately large values of $\tau$.
The prominent deviations of some experimental results from the FES predictions are likely due to small fabrication defects in the samples, and possibly residual pre-tension.
In addition, as $\tau$ increases, the deviation of each markers in our FEM data becomes prominent [see the insets of Fig~\ref{Figure09}~(b) and (c)].
This trend likely reflects the limitations of the present first-order-in-$\tau$ description (see SM~\S~IV).

%%% FIGURE 8 

%%% FIGURE 9

\subsection{Relative frictional stress}

The dependence of $m$ and $d_w/h$ solely on $\tau$ reflects the fact that our problem is fully characterized by the two dimensionless parameters $d/h$ and $\tau$~\cite{box2017indentation}.
To make this explicit, we choose a non-dimensionalization that does not contain $d$: 
\begin{align}
    \tilde{r} \equiv \frac{r}{\ell_g},~~ \tilde{w}(\tilde{r}) \equiv\frac{w}{h}, ~~ \tilde{\sigma}_{\alpha\beta}(\tilde{r}) \equiv \frac{\sigma_{\alpha\beta}}{B/(h\ell_g^2)}, ~~ \tilde{K}_{\alpha\beta}(\tilde{r}) = \frac{{K}_{\alpha\beta}}{h/\ell_g^2}.
    \label{sec06C:nondim}
\end{align}
These scaling factors can also be obtained from Eq.~(\ref{sec04A:def:nondim}) or~(\ref{sec04B:def:nondim}) with $d=h$.
With this non-dimensionalization, we find that Eq.~(\ref{sec04:eq:FvKout}) becomes
\begin{align}
    \tilde{\nabla}^4 \tilde{w} -  (\tilde{K}_{rr}\tilde{\sigma}_{rr} + \tilde{K}_{\theta\theta}\tilde{\sigma}_{\theta\theta}) +1 = 0. \label{sec06C:FvKout}
\end{align}
It can be seen that $\ell_g$ is the essential length scale in this problem, since normalizing all lengths by $\ell_g$ scales out the FvK equations.

In the absence of friction, since Eq.~(\ref{sec06C:FvKout}) does not contain any parameters; only the boundary condition,
\begin{align}
\tilde{w}(0) = \frac{d}{h}
\end{align}
characterizes the physical behavior of the sheet.
We thus expect $d_w/h\sim O(1)$ at the instability, and consequently $m\sim (d_w/h)^2\sim O(1)$.

In the frictional case, we focus the compressive hoop stress term in Eq.~(\ref{sec06C:FvKout}), as it is most influenced by the friction.
Using Eqs.~(\ref{sec05B:eq:srr}) and (\ref{sec05B:eq:stt}), we find
\begin{align}
\tilde{\sigma}_{\theta\theta} \sim \frac{-\sigma_{rr} + \mu \rho g a}{B/(h\ell_g^2)} = -\tilde{\sigma}_{rr}+\tau,
\label{sec06C:stt}
\end{align}
which contains an additional parameter, $\tau$.
Accordingly, the problem is governed by two parameters, $(d/h,\tau)$.

The relative frictional stress $\tau$ is reminiscent of the parameter referred to as ``mechanical bendability'' in the previous studies \cite{davidovitch2011prototypical, vella2018regimes, box2017indentation}.
Note, however, that while the mechanical bendability in these works is defined as the ratio of the radial tension to the bending force, $\tau$ instead quantifies the frictional contribution to the hoop stress. 
In the problem considered in the above references, the bendability is so large that the physically relevant regime is \textit{far from threshold}, where the stress state is well described by \textit{tension-field theory}~\cite{mansfield1989bending, davidovitch2011prototypical, vella2019buffering}. 
In that limit, the hoop stress is asymptotically small compared with the radial stress, i.e., $|\sigma_{\theta\theta}|/\sigma_{rr}\to 0$.
In contrast, our study focuses on a moderately bendable regime, in which $\sigma_{rr}$ and $\sigma_{\theta\theta}$ remain of the same order, and the contribution of $\tau$ to $\sigma_{\theta\theta}$ in Eq.~(\ref{sec06C:stt}) is subdominant.
Therefore, our problem lies in a \textit{near-threshold} regime, distinct from the highly bendable limit, and is more closely related to problems of wrinkling in the indentation of moderately thick floating sheets~\cite{box2017indentation} and in liquid blister tests~\cite{chopin2008liquid}.

%%% FIGURE 10
\begin{figure*}[]
    \centering	\includegraphics[width=.75\linewidth]{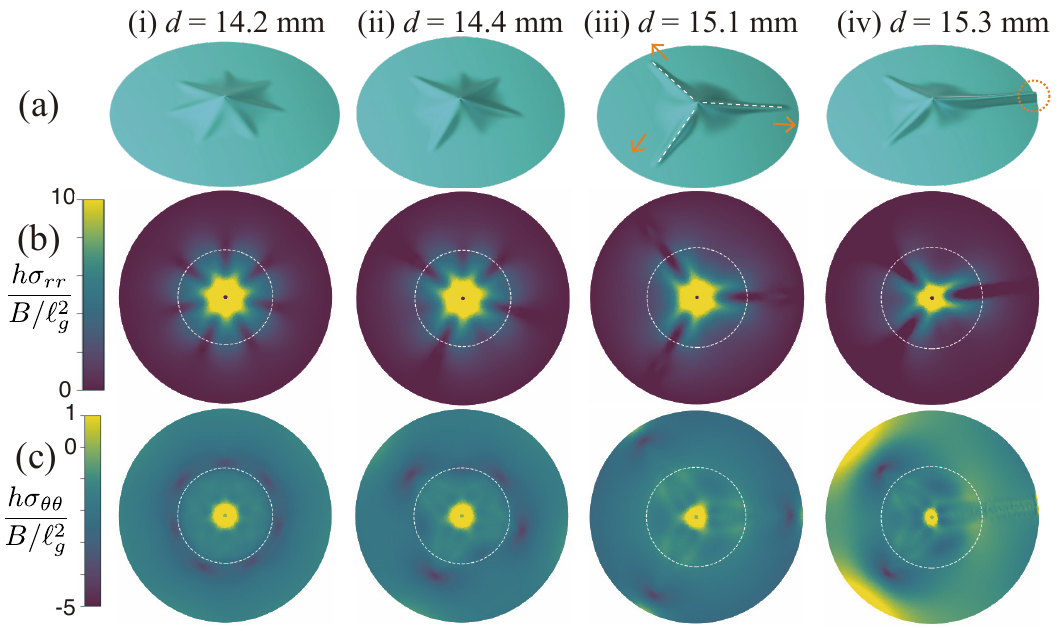} 	\caption{
    Evolution of the deformation from (i) the seven-fold symmetric wrinkle pattern
to (ii) a distorted, less-ordered pattern, (iii) a triangular-like shape, and finally (iv) the post-buckling configuration.
    (a) Shape transition.
    (b, c) Radial and azimuthal components of the stress distribution.
    The color indicates the value of dimensionless stress.
    The dotted circle denotes the radius $R$ calculated by Eqs.(\ref{sec03:eq:R}) and~(\ref{sec04B:sol:cs}).
    All results are obtained from FES with parameters $(E, h, a, \mu) = (226~\textrm{kPa}, 0.5~\textrm{mm}, 150~\textrm{mm}, 0)$.
 }\label{Figure10}
\end{figure*}

\section{Global buckling} ~\label{sec:globalbuckling}

In this section, we focus on the instabilities induced by large indentation, in particular the morphological transitions leading to global buckling.

In the post-wrinkled state, the lifted region continues to grow with the indentation height $d$, as described in Eq.~(\ref{sec03:eq:R}).
As shown in Fig.~\ref{Figure10}~(a), the initially well-ordered $m$-fold wrinkles become distorted as $d$ increases.
Eventually, three to four dominant wrinkles grow preferentially, while the others disappear, resulting in a polygonal lifted shape with typically $m=3-4$.
The tips of these dominant wrinkles extend to the end of the sheet, as indicated by the orange arrow in Fig.~\ref{Figure10}~(a-iii).
When the hoop stress at $r\sim a$, which is estimated as $|\sigma_{\theta\theta}(a)|\sim E|\epsilon_{\theta\theta}(a)|\sim E(d/a)^2$, exceeds a critical threshold, localized buckling occurs near the sheet’s periphery~\cite{chopin2008liquid}, as shown in Fig.~\ref{Figure10}.
This peripheral buckling is accompanied by discontinuous changes in force and stress~\cite{holmes2010draping}, as seen in Fig.~\ref{Figure02} and~\ref{Figure10}.
In both experiments and FES, we find that multiple wrinkle tips sometimes buckle simultaneously, whereas in other cases a single ruck forms first (Fig.~\ref{Figure10}).

\begin{figure}[]
    \centering	\includegraphics[width=1.\linewidth]{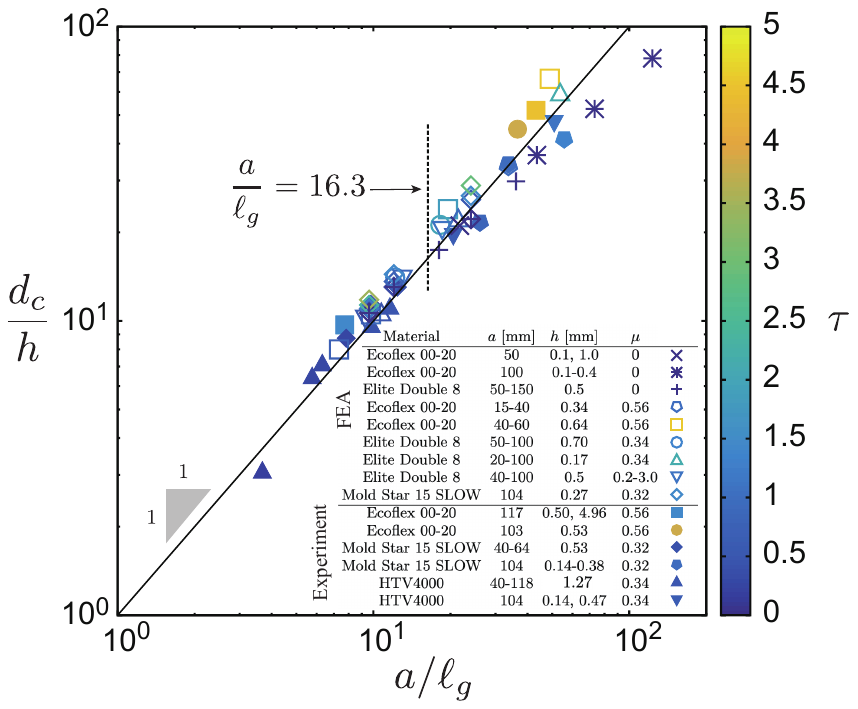} 	\caption{Critical displacement of global buckling. 
    The results obtained from FES and experiments are represented by open and filled symbols, respectively.
    The markers $\times$, $\ast$, and $+$ indicate the results from FES without friction.
    The color bar indicates the magnitude of $\tau$.
    The solid line represents $d_c/h=a/\ell_g$, which corresponds to Eq.~(\ref{sec07:eq:dc}).
    The dashed line denotes the value of $a/\ell_g = 16.3$ given in Eq.~(\ref{sec07A:ineq}).
 }\label{Figure11}
\end{figure}

Here, we assume that the energy of each dominant wrinkle is nearly identical before global buckling, and estimate the energy per unit area of a single ruck formed at the periphery as:
\begin{align}
    e\sim B \frac{A^2}{\lambda^4} - h |\sigma_{\theta\theta}| \frac{A^2}{\lambda^2} + \rho g h A,  \label{sec07:energy:ruck}
\end{align}
where $A$ and $\lambda$ are the amplitude and width of the ruck, respectively.
Here, we take the pre-buckling state as the reference, with $e_{\rm pre}=0$, corresponding to $A=0$.
Assuming conservation of the arc-length along the azimuthal direction at $r=a$, $A$ is related to $\lambda$ via $A\sim |\epsilon_{\theta\theta}|^{1/2} \lambda$; therefore $e$ is a function of the $\lambda$ only.
By minimizing $e$ with respect to $\lambda$, the characteristic wave length of the ruck is obtained as follows:
\begin{align}
    \lambda \sim \ell_{gb}|\epsilon_{\theta\theta}|^{1/6}, \label{sec07:eq:lambda}
\end{align}
where $\ell_{gb}\equiv (B/\rho g h)^{1/3}$ is the gravito-bending length~\cite{wang1986critical}, which is distinct from $\ell_g=(B/\rho g)^{1/4}$ as noted in Section 3.
The minimized energy for the case $A\neq 0$ is obtained substituting Eq.~(\ref{sec07:eq:lambda}) into Eq.~(\ref{sec07:energy:ruck}) as
\begin{align}
    e_{\rm buckled}\sim \frac{B}{\ell_{gb}^2} |\epsilon_{\theta\theta}|^{2/3} - Eh|\epsilon_{\theta\theta}|^2. \label{sec07:energy:buckled}
\end{align}
The buckling threshold is determined by the requirement that the formation of a ruck reduces the energy: $e_{\rm buckled}< e_{\rm pre}$.
This condition can be summarized as $d_c \sim a ({\rho g h}/{E})^{1/4}$, or in dimensionless form:
\begin{align}
    \frac{d_c}{h} \sim  \frac{a}{\ell_g}.  \label{sec07:eq:dc}
\end{align}
We plot $d_c/h$ as obtained from the experiments and FES as a function of $a/\ell_g$ in Fig.~\ref{Figure11}.
As predicted, $d_c/h$ and $a/\ell_g$ exhibit an approximately proportional relationship, with a coefficient nearly equal to 1.

%%% FIGURE 11

\subsection{Few-mode wrinkle formation driven by global buckling}

When various sheet parameters are fixed, and only the thickness is increased, or the radius is decreased, the critical displacement for global buckling $d_c$, eventually becomes smaller than the wrinkle-onset threshold $d_w$.
The condition $d_w > d_c$ appears to suggest that global buckling can occur directly from the axisymmetric state without the intermediate formation of wrinkles.  
Neglecting friction for simplicity, and using the numerical coefficients indicated by the dotted lines in Fig.~\ref{Figure07} and Fig.~\ref{Figure11}, the condition $d_w > d_c$ can be written as
\begin{align}
    a \lesssim 16.3 \, \ell_g. \label{sec07A:ineq}
\end{align}

\begin{figure}[]
    \centering	\includegraphics[width=1.\linewidth]{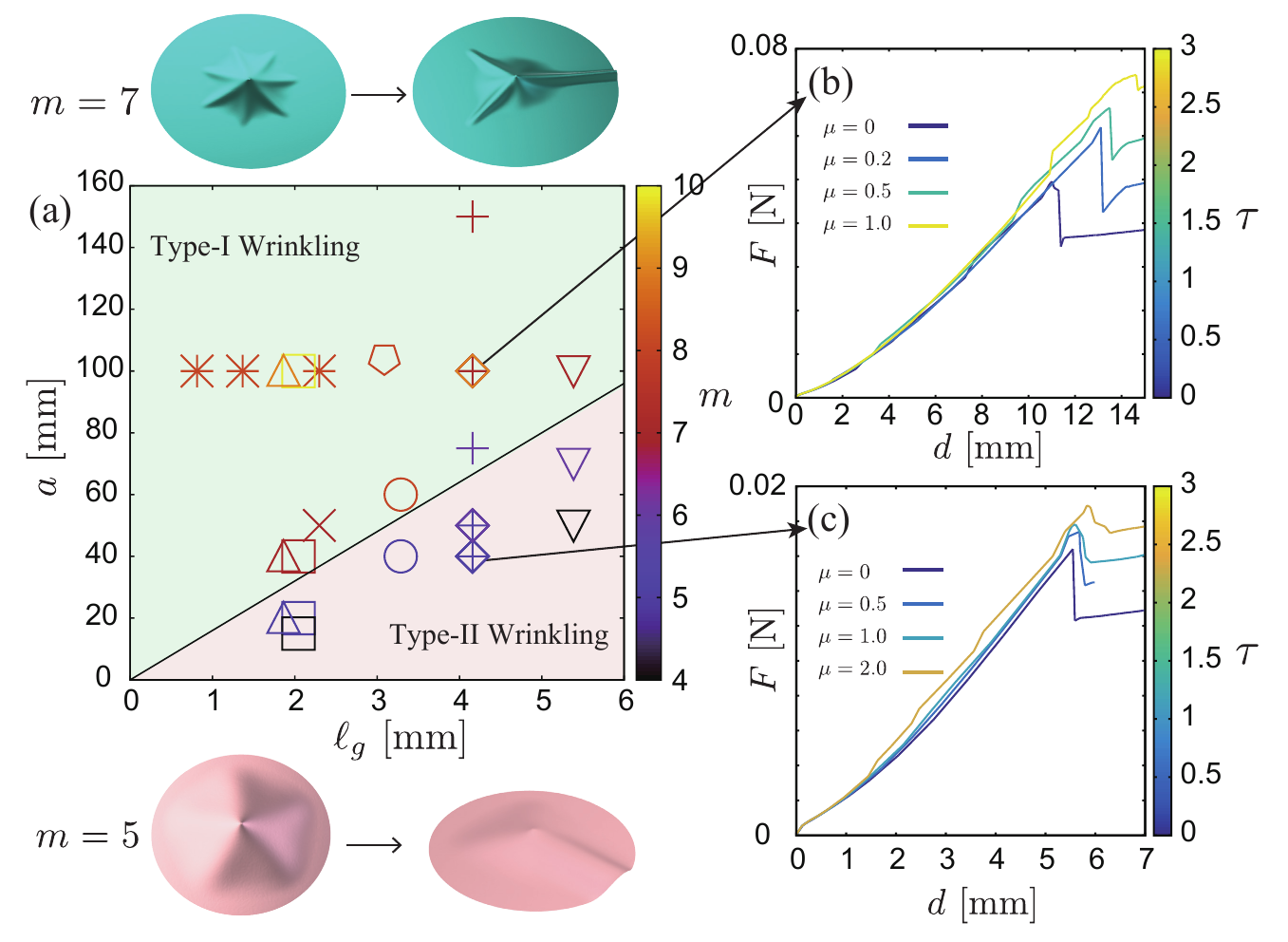} 	\caption{
    (a) Phase diagram classifying the observed wrinkling patterns into Type-I (with seven or more folds) and Type-II (with fewer folds).
    Details of the symbols are summarized in the inset of Fig.~\ref{Figure11}.
    The solid line indicates the equality condition in~(\ref{sec07A:ineq}).
    The CG illustrations generated from FES show the transition of the shape from a wrinkled to a globally buckled state.    (b, c) Force--displacement curves obtained from finite-element simulations with varying $\mu$, plotted for two cases:
    (b) a transition from Type-I wrinkling to global buckling, and (c) a transition from Type-II wrinkling state to global buckling.
 }\label{Figure12}
\end{figure}

The colormap in Fig.~\ref{Figure12} shows the number of wrinkles obtained from FES at the earliest onset of instability prior to global buckling.
Figure~\ref{Figure12} (a) indicates that Eq.~(\ref{sec07A:ineq}) acts as a boundary that separates the phases with fewer (possibly only transient) wrinkles (typically $m = 5$) and those with statically formed wrinkles ($m \ge 7$). 
We emphasize that this trend is consistent with the behavior observed in Fig.~\ref{Figure07} for the range $a/\ell_g \lesssim 20$.
We classify the wrinkling observed in the regime $a/\ell_g \gtrsim 16.3$---which has been the focus of this study---as \textit{Type-I Wrinkling}, and define the other regime as \textit{Type-II Wrinkling}.
Type-II wrinkles were observed only briefly, immediately before the onset of global buckling, and disappeared as soon as the periphery of the sheet buckled.  
Therefore, Type-II wrinkles are presumed to form as a transient feature during the direct transition from the axisymmetric state to the globally buckled state, and are likely governed by a mechanism distinct from that of the Type-I wrinkles discussed in \S~\ref{sec:wrinkling}.

As the sheet becomes even thicker and/or its radius decreases, physically different behaviors may also emerge, including a simple case in which the sheet is trivially lifted off without any buckling at all.  
A detailed investigation of these phenomena, including the transient Type-II wrinkles, is beyond the scope of the present study, and is left for future works.

%%% FIGURE 12

\subsection{Dependence of friction}

In the preceding argument, which leads to Eq.~(\ref{sec07:eq:dc}), we neglected the contribution of friction. However, a weak dependence on the friction coefficient can be observed in Fig.~\ref{Figure11}.
For each $\tau$, the deviations from Eq.~(\ref{sec07:eq:dc}) in both the experiments and FES exhibit qualitatively similar trends.

Figures~\ref{Figure12}~(b) and (c) show the force--displacement curves obtained for sheets in the Type-I and Type-II phases, respectively, for different values of $\mu$.
We clearly see that while $d_c$ in Type I is significantly affected by friction, $d_c$ in Type II is rather insensitive to $\mu$.
This might be reasonable, considering that, whereas Type-II wrinkle is the only transient pattern appearing just before the global buckling, Type-I wrinkle is a static stable structure formed at the preceding stage of the global buckling instability. 
During the emergence of the Type-I wrinkles, the contact area and geometry of the sheet on the frictional surface may become highly complicated, which could substantially influence the onset of the global buckling.
Indeed, as shown in Fig.~\ref{Figure11}, the data exhibit scatter for $a/\ell_g > 16.3$, where the deviation of $d_c/h$ from the scaling prediction is larger for larger values of $\tau$.

%%% FIGURE 13
\begin{figure*}[t]
    \centering	\includegraphics[width=1.\linewidth]{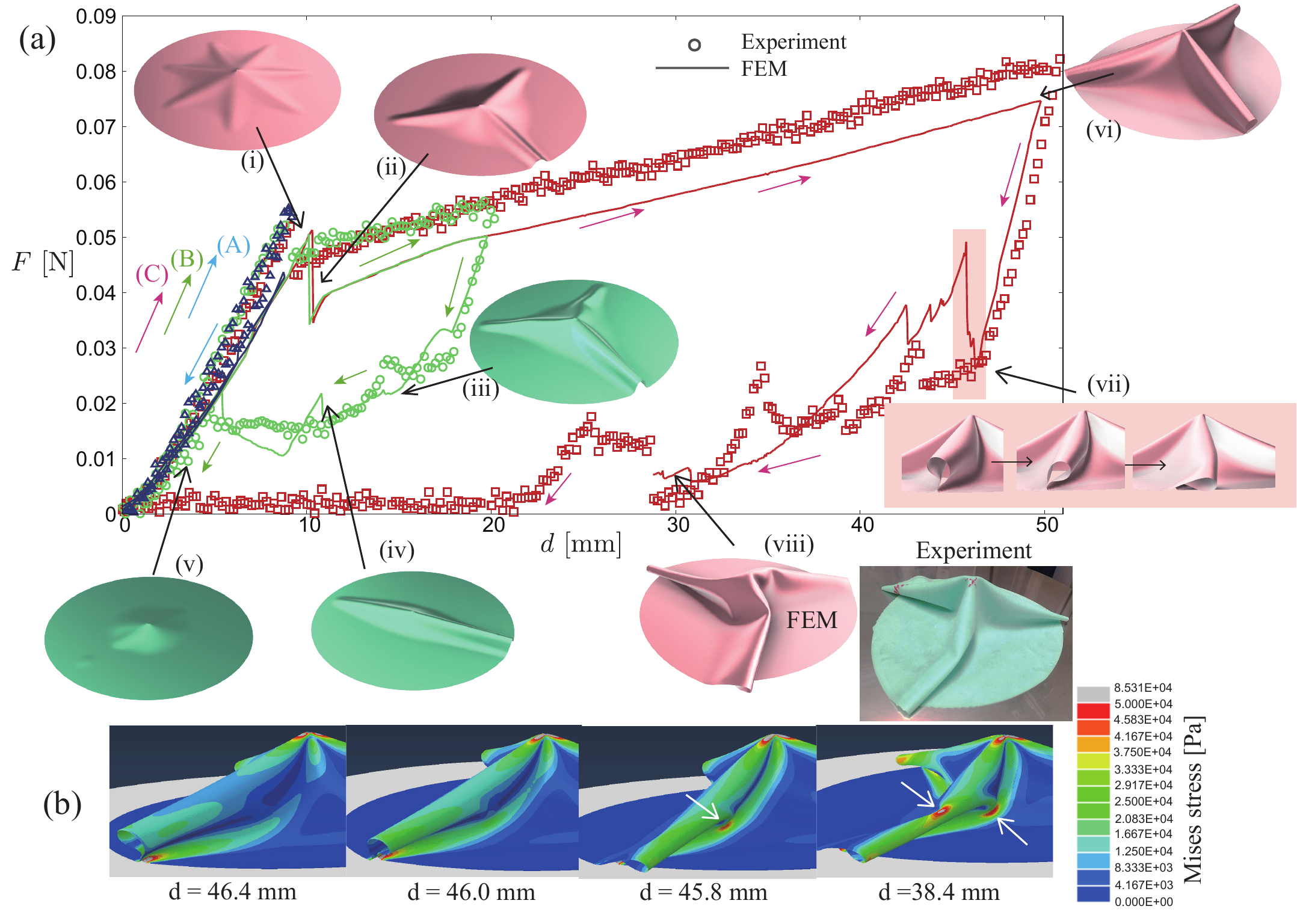} 	\caption{
    (a) Force--displacement curve for the one cyclic test.
    We changed the maximum indentation height, $d_{\rm max}$.
    The blue, green, and red colors represent the indentation forces for $d_{\rm max}<d_c$, $d_{\rm max}>d_c$, and $d_{\rm max}\gg d_c$, respectively.
    The open symbols and solid lines represent experimental and numerical data, respectively.
    The illustrations of sheets are obtained from FES.
    (b) von Mises stress distribution in the inset (vii) of panel (a), calculated in Abaqus. 
    Stress focusing is observed in some areas.
    All results are obtained from the experiment and FES with parameters $(E, h, a, \mu) = (477~\textrm{kPa}, 0.38~\textrm{mm}, 104~\textrm{mm}, 0.32)$.
 }\label{Figure13}
\end{figure*}

\section{Hysteresis}

Beyond the global buckling, the shape of the sheet undergoes large geometrically nonlinear deformations, self-contact, and folding.
Figure~\ref{Figure13} presents the experimental and FES data for a protocol in which $d$ is increased and then reduced to $d=0$, in a quasi-static manner.
In such a cyclic indentation test, we observe dramatic shape changes, including ruck formation and folding.
Whether these remain permanent after the load is removed depends on the maximum indentation height, $d_{\rm max}$. 
The morphological process of the buckled sheet is closely related to the magnitude of hysteresis observed in the force curves in Fig.~\ref{Figure13}.
Specifically, we performed both experiments and FES to investigate the following three cases: (A) $d_{\rm max}<d_c$; (B) $d_{\rm max}$ is slightly larger than $d_c$; and (C) $d_{\rm max}\gg d_c$.
Below, we describe the three qualitatively distinct behaviors of the sheet observed in the experiment and FES.

For case (A), the global buckling of the sheet is absent because $d_{\rm max}<d_c$, and no appreciable hysteresis in the force curve is observed, confirming that the cycle is reversible even in the presence of the Type-I wrinkles.

For case (B), a sheet is deformed beyond the global buckling point, and its corresponding force curve exhibits pronounced hysteresis.
As $d$ increases, significant stretching is stored in the pre-buckled state [see Fig.~\ref{Figure13}~(a-i)], and this in-plane strain is released abruptly at the onset of global buckling [see Fig.~\ref{Figure10} and \ref{Figure13}~(ii)]~\cite{chopin2008liquid, holmes2010draping, suzanne2022indentation, montalvo2023wrinkling}.
Because the stress state in the sheet is largely different, the shape change behaviors can naturally differ between the ascending and descending processes~(iii--iv).
However, in this regime, the rucks subsequently slip and unfold as $d$ is reduced~\cite{vella2009statics} such that no rucks or folds remain once $d$ becomes sufficiently small. 
The force--displacement curve then gets back to the same path as that in the increasing process~(v)~\cite{holmes2010draping}, and at $d=0$ the sheet returns to its initial flat configuration.
In this sense, the cycle process in (B) may be characterized by a single closed loop in the configurational space. 

In contrast, for case (C), the sheet is lifted well beyond the global buckling~($d_{\rm max}\gg d_c$, and (vi) in Fig.~\ref{Figure13}~(a)), we observe the rucks remain even when $d$ decreases to zero~\cite{vella2009statics}.
The rucks formed are initially symmetric, but they often collapse to one side, either the left or right (by spontaneous symmetry breaking), as previously observed in heavy elastica~\cite{wang1986critical, domokos2003symmetry}, leading to a partially folded sheet configuration. 
Such a self-folded ruck exhibits a stress focused structure \cite{witten2007stress} as shown in Fig.~\ref{Figure13}~(b).
Moreover, these behaviors correspond to a seemingly mysterious discontinuous increase in indentation force $F$ during the descent process~(vii).
Some of the residual rucks and folds prevent the sheet from returning to its initial flat configuration and remain permanently owing to the coupling among the intricate 3D sheet geometry, self-contacts, and frictional interactions.
The full sequence is also shown in SM Movie, where the experiment and the FES are shown together for case (C) in Fig.~\ref{Figure13}.

Although these are the typical behaviors in cases (A)--(C),  the sheet can behave in more complicated ways depending on the magnitude of friction. 
For example, unfolding events such as those observed in case (B) occur only when the surface friction is sufficiently weak.
In addition, folds produced by collapsed rucks may either slide along the sheet surface while remaining in self-contact~\cite{demery2014mechanics} or lift back up into rucks during unloading.
Furthermore, for large values of $\tau$, irreversible shape change of the sheet can be observed even for $d_{\rm max} < d_c$.
Understanding and quantitatively classifying such morphological diversity is far from straightforward, and further investigations will be needed to uncover the underlying mechanisms.

\section{Conclusion}\label{sec:conclusion}

In this study, we combined analytical theory, experiments, and numerical simulations to investigate the indentation response of a heavy elastic sheet resting on a frictional substrate.  
We uncovered the fundamental principles governing pattern formation under indentation, including nonlinear and discontinuous force responses, wrinkling instabilities, and global buckling.
A characteristic length scale, $\ell_g$, was identified, which not only controls the size of the uplifted region and the force response but also emerges consistently in the onset of wrinkling and the transition to global buckling.  
In the frictionless case, we discovered a remarkably simple rule: the number of wrinkles formed is constant $(m\approx 7)$ and the critical indentation height depends solely on the sheet thickness ($d_w\approx 16.3 h$). 
When friction is present, the wrinkling behavior becomes more complex; however, it can still be characterized in terms of another dimensionless parameter, the relative frictional stress $\tau$. 
Together, these results provide a quantitative framework for the indentation-induced morphology of heavy sheets on frictional substrates.

One might suppose that the post-buckled shapes observed in our experiment resemble the developable configurations previously studied for suspended sheets under gravity~\cite{cerda2004elements}.
However, these systems differ fundamentally in their internal stresses and relevant boundary conditions.
Obviously, a draped sheet is force- and torque-free at its outmost edge.
In contrast, because the globally buckled sheet studied here is still partially in contact with the substrate, it is generally subjected to more complicated boundary conditions at its edge owing to the residual in-plane strains within the sheet. 
Indeed, we have frequently observed a discontinuous transition to a developable surface in our experiment, which occurs when the outmost edge of a buckled sheet fully detaches from the substrate.
The resulting developable configuration again contains wrinkles, but their number generally does not coincide with that in the globally buckled phase. 
This difference likely explains the discontinuity of the transition, although further investigations will be needed.

Thus far, we have discussed only the near-threshold behavior of the wrinkling instability.
As shown in Fig.~\ref{Figure10}, well-developed wrinkles become distorted, with only a few dominant wrinkles growing preferentially.
The linear stability analysis cannot explain this collapse or the process by which the number of dominant wrinkles is determined. 
Furthermore, our findings in Eqs.~(\ref{sec06B:eq:m_fric}) and~(\ref{sec06B:eq:dw_fric}) apply only in the regime of relatively weak friction.
As seen in Fig.~\ref{Figure09}, when $\tau \gtrsim 10$, $m$ and $d_w/h$ are no longer characterized solely by $\tau$.
The materials and sheet sizes used in our experiments do not reach $\tau>10$ for typical friction coefficients.
However, if we are to apply our study to extremely large systems, such as those in aerospace engineering~\cite{spencer2019solar} or geophysical contexts~\cite{turcotte2002geodynamics}, our theory should be extended to account for the high-$\tau$ regime, since $\tau$ is proportional to system size.

Finally, despite the extreme simplicity of the loading protocol, the sheet exhibits a remarkably rich set of configurations ranging from axisymmetric uplifts and wrinkle patterns to global buckling.
Upon unloading from sufficiently large lifts, we observed a folded configurations stabilized by self-contact and friction, as shown in Fig.~\ref{Figure13} and in the ``ghost'' in Fig.~\ref{Figure01}(a).
The resulting structures can be viewed as gravity-driven \textit{self-organizing origami}~\cite{mahadevan2005self, felton2014method, kim2025construction}.
Our findings may contribute to the controlled design of well-ordered wrinkles or artistic complex folds through contact interactions under gravity, fluid drag, or high pressure.

\section*{Author contributions}
\textbf{Keisuke Yoshida}: Conceptualization, Data curation, Formal analysis, Funding acquisition, Investigation, Methodology, Software, Validation, Visualization, Writing – original draft, Writing – review \& editing.
\textbf{Hirofumi Wada}: Conceptualization, Formal analysis, Funding acquisition, Project administration, Resources, Supervision, Validation, Writing – review \& editing.

\section*{Conflicts of interest}
There are no conflicts of interest to declare.

\section*{Data availability}
The data that support the findings of this article are available within the article and within its Supplemental Material.

\section*{Acknowledgements}
We acknowledge financial support from JSPS KAKENHI (Grant No. 23K22463 to H.W.); a Grant-in-Aid for JSPS Research Fellows (DC1; Grant No. 21J22837 to K.Y.); the Early-career Researcher Development Program of Ritsumeikan University (to K.Y.); and a research grant of 2025 (during 2025-2026) from Amano Institute of Technology (K.Y.).

ChatGPT (OpenAI) was used for language editing and minor assistance with data-processing scripts. The authors take full responsibility for the content and results.

\bibliography{bibliography}

\clearpage


\section*{Supplementary material}

# Supplemental Material:
# Wrinkles, rucks and folds formed in a heavy sheet on a frictional surface

Keisuke Yoshida[1,2][*] and Hirofumi Wada[1]

[1]*Department of Physical Sciences, Ritsumeikan University, Kusatsu, Shiga 525-8577, Japan and*
[2]*Research Organization of Science and Technology,*
*Ritsumeikan University, Kusatsu, Shiga 525-8577, Japan*

(Dated: January 1, 2026)

## I. EFFECT OF INDENTER SIZE

To examine the effect of indenter radius on the load–displacement and lifted-radius–displacement relations, we performed finite-element simulations (FES). We chose a circular elastic sheet with $(E, \nu, \rho, h, a, \mu) = (477 \text{ kPa}, 0.47, 1127 \text{ kg/m}^3, 0.27 \text{ mm}, 100 \text{ mm}, 0.31)$, for which $\ell_g = (B/\rho g)^{1/4} \approx 3.1$ mm. Indentation was imposed by gradually increasing a vertical displacement $d$ over a central circular region $0 \leq r \leq r_{\text{ind}}$. We varied the indenter radius as $r_{\text{ind}} = 0.1, 0.5, 1, 2$, and 5 mm. As shown in Fig. S1, the numerical data approach the point-load predictions in Eqs. (16) and (17) as $r_{\text{ind}}$ decreases. In the moderate deformation regime $(d/h \sim 10^{-1})$, when $r_{\text{ind}} \ll \ell_g(d/h)^{1/4}$, the indentation can be regarded as a concentrated load acting on the sheet. By contrast, for $d/h \gg 1$ the characteristic lifted radius $\ell_g(d/h)^{3/4}$ is typically much larger than $r_{\text{ind}}$ in our setup, so the point-load approximation is well satisfied.

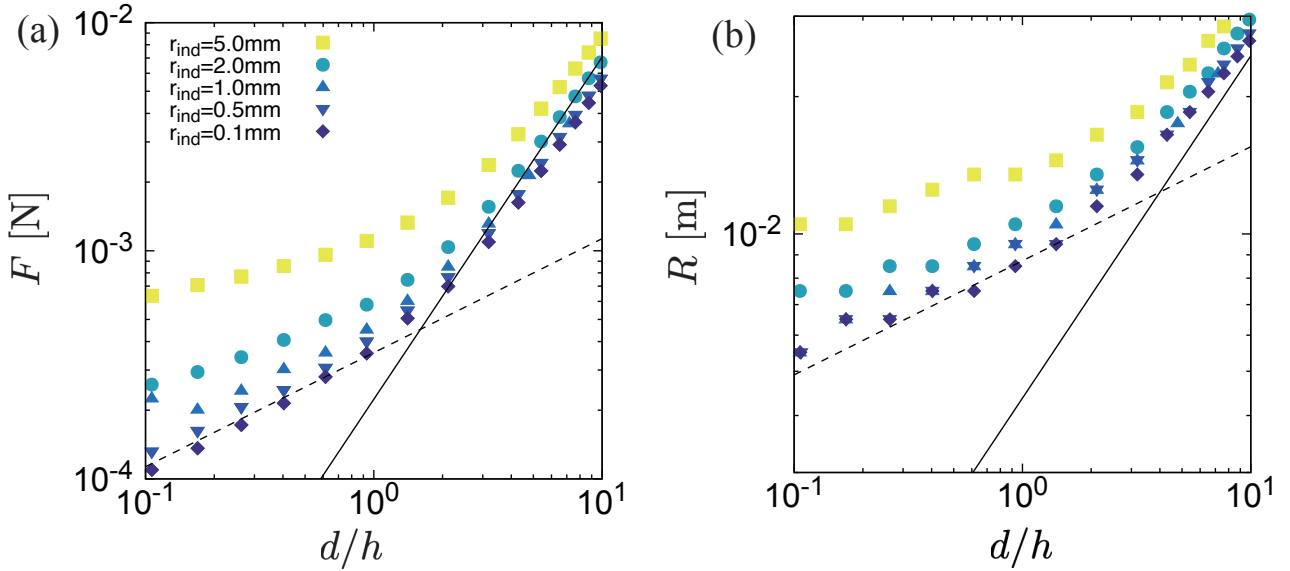

FIG. S1. FES results for varying indenter radii $r_{\text{ind}}$ (all other parameters identical). (a) Load–displacement relation. (b) Lifted radius–displacement relation. The dashed and solid lines represent Eq. (16), (17), (23), and (24) in the main text, which are derived under the point-loading assumption. In both cases, as $r_{\text{ind}}$ decreases, the numerical results approach the theoretical predictions based on the point-load model.

---

[*] kyosh424@gmail.com

## II. FULL DERIVATION OF THE MEMBRANE SOLUTIONS

When the indentation height is sufficiently larger than the sheet thickness ($d/h \gg 1$), the equilibrium of a thin membrane subjected to a concentrated load $\mathcal{F}$ is described by the coupled differential equations for the out-of-plane displacement $W(\xi)$ and the Airy stress function $\Psi(\xi)$:

$$-\Psi W' + \frac{\alpha}{2}\xi^2 - \frac{\mathcal{F}}{2\pi} = 0, \tag{S1}$$

$$\xi \frac{d}{d\xi}\left[\frac{1}{\xi}\frac{d}{d\xi}(\xi\Psi)\right] + \frac{1}{2}W'^2 = 0, \tag{S2}$$

where $\xi$ denotes the radial coordinate. The quantities $\xi$, $W$, and $\Psi$ are nondimensionalized according to Eq. (18) in the main text.

Equations (S1) and (S2) with the boundary conditions

$$W(0) = 1, \; W(1) = 0, \; \Psi(0) = 0, \; \Psi'(1) - \nu\Psi(1) = 0, \tag{S3}$$

can be solved analytically in the absence of gravity ($\alpha = 0$) [1]. In particular, for $\nu = 1/3$, the so-called *Schwerin's solution* [2, 3] is given by

$$\Psi^{(0)} = \frac{1}{4}\xi^{1/3}, \tag{S4}$$

$$W^{(0)} = 1 - \xi^{2/3}, \tag{S5}$$

$$\mathcal{F}^{(0)} = \frac{\pi}{3}. \tag{S6}$$

We now consider the case including gravity. The effect of gravity, characterized by the dimensionless parameter $\alpha$ in (S1), is assumed to be small enough to be treated as a perturbation. Expanding $\Psi$, $W$, and $\mathcal{F}$ around the unperturbed solutions:

$$\Psi = \Psi^{(0)} + \alpha\Psi^{(1)} + O(\alpha^2), \tag{S7}$$

$$W = W^{(0)} + \alpha W^{(1)} + O(\alpha^2), \tag{S8}$$

$$\mathcal{F} = \mathcal{F}^{(0)} + \alpha\mathcal{F}^{(1)} + O(\alpha^2). \tag{S9}$$

Solving Eqs. (S1) and (S2) up to $O(\alpha)$ yields

$$W(\xi) = 1 - \xi^{2/3} + \alpha\left(\frac{5}{4}\xi^{8/3} - \frac{1}{4}\xi^{2/3} - \xi^2\right), \tag{S10}$$

$$\Psi(\xi) = \frac{1}{4}\xi^{1/3} + \alpha\left(\frac{1}{2}\xi^{7/3} + \frac{1}{8}\xi^{1/3} - \frac{3}{4}\xi^{5/3}\right), \tag{S11}$$

$$\mathcal{F} = \frac{\pi}{3} + \alpha\frac{\pi}{4}. \tag{S12}$$

The stress components follow from the derivatives of $\Psi$:

$$\Sigma_{rr} \equiv \frac{\Psi}{\xi} = \frac{1}{4}\xi^{-2/3} + \alpha\left(\frac{1}{2}\xi^{4/3} + \frac{1}{8}\xi^{-2/3} - \frac{3}{4}\xi^{2/3}\right), \tag{S13}$$

$$\Sigma_{\theta\theta} \equiv \frac{d\Psi}{d\xi} = \frac{1}{12}\xi^{-2/3} + \alpha\left(\frac{7}{6}\xi^{4/3} + \frac{1}{24}\xi^{-2/3} - \frac{5}{4}\xi^{2/3}\right). \tag{S14}$$

The dimensionless stresses $\Sigma_{\alpha\beta}$ are related to $\sigma_{\alpha\beta}$ through

$$\sigma_{\alpha\beta} = \frac{Ed^2}{R^2}\Sigma_{\alpha\beta}. \tag{S15}$$

To determine the undetermined parameter $\alpha$, we estimate it by minimizing the total energy [3]. First, the stretching

energy is obtained from Eqs. (S10-S14):

$$\mathcal{E}_s = \frac{h}{2} \int_A \sigma_{\alpha\beta} \epsilon_{\alpha\beta} dA \tag{S16}$$

$$= \frac{Ehd^4}{2R^2} \oint d\theta \int_0^1 \xi d\xi \left\{ (\Sigma_{rr} + \Sigma_{\theta\theta})^2 - 2(1+\nu)\Sigma_{rr}\Sigma_{\theta\theta} \right\} \tag{S17}$$

$$= \frac{\pi Ehd^4}{R^2} \left\{ \frac{5-3\nu}{48} + \alpha \frac{\nu - 1/3}{16} + O(\alpha^2) \right\}. \tag{S18}$$

For $\nu = 1/3$, this simplifies to

$$\mathcal{E}_s = \frac{\pi Ehd^4}{R^2} \left\{ \frac{1}{12} + O(\alpha^2) \right\}. \tag{S19}$$

The gravitational energy is given by

$$\mathcal{E}_g = \rho g h \int_A w \, dA \tag{S20}$$

$$= \rho g h R^2 d \oint d\theta \int_0^1 \xi W(\xi) d\xi \tag{S21}$$

$$= 2\pi \rho g h R^2 d \left( \frac{1}{8} - \alpha \frac{17}{224} + O(\alpha^2) \right) \tag{S22}$$

Minimizing $\mathcal{E}_s + \mathcal{E}_g$ with respect to $R$ at the leading order ($O(\alpha^0)$) yields

$$R \approx 3^{-1/4} \left( \frac{E}{\rho g} \right)^{1/4} d^{3/4} \tag{S23}$$

$$= [4(1-\nu^2)]^{1/4} \ell_g \left( \frac{d}{h} \right)^{3/4}. \tag{S24}$$

Comparing Eq. (S24) with Eq. (1) in the main text, we identify

$$c_s = [4(1-\nu^2)]^{1/4} = \left( \frac{32}{9} \right)^{1/4}, \tag{S25}$$

where we have substituted $\nu = 1/3$ in accordance with the above assumption. By definition, once $c_s$ is determined we obtain

$$\alpha = \frac{c_s^4}{12(1-\nu^2)} = \frac{1}{3}. \tag{S26}$$

Using Eq. (S9), the corresponding load becomes $\mathcal{F} \approx 5\pi/12$, and hence

$$F \approx \frac{Ehd^3}{R^2} \mathcal{F} \tag{S27}$$

$$\approx \frac{5\pi(1-\nu^2)}{c_s^2} \rho g h \ell_g^2 \left( \frac{d}{h} \right)^{3/2}. \tag{S28}$$

Thus, the prefactor $k_s$ in the scaling laws for the force response $F$ [Eqs. (3) in the main text] is

$$k_s \approx \frac{5\pi}{2} \sqrt{1-\nu^2} = \frac{5\sqrt{2}\pi}{3}. \tag{S29}$$

## A. Rescaling of nondimensional stress

In the main text, the stress distribution is plotted in units of $B/(h\ell_g^2)$, which serves as the characteristic stress scale independent of $d$. Here, we express $\Sigma_{\alpha\beta}$ in terms of $h\ell_g^2 \sigma_{\alpha\beta}/B$ by using Eqs. (S15) and (S24):

$$\frac{h\sigma_{\alpha\beta}}{B/\ell_g^2} = \frac{Ehd^2\ell_g^2}{BR^2}\Sigma_{\alpha\beta} \tag{S30}$$

$$= \frac{12(1-\nu^2)}{c_s^2}\left(\frac{d}{h}\right)^{1/2}\Sigma_{\alpha\beta} \tag{S31}$$

$$\approx 4\sqrt{2}\left(\frac{d}{h}\right)^{1/2}\Sigma_{\alpha\beta}. \tag{S32}$$

## III. ESTIMATION OF THE STICK–SLIDE BOUNDARY

We focus on the planar region of the sheet ($R < r < a$). A frictional force per unit area, $\boldsymbol{f}(r,\theta) = f_r \hat{\boldsymbol{e}}_r + f_\theta \hat{\boldsymbol{e}}_\theta$ acts between the sheet and the substrate. As the indentation height $d$ is gradually increased, the in-plane stress in the sheet may locally exceed the maximum static friction, so that regions of sticking and sliding can coexist.

Here we consider the case in which $R < r < r_*$ is the sliding region, while $r_* < r < a$ corresponds to the sticking region. For $r \gg r_*$, the traction remains smaller than the maximum static friction, so material points stay pinned to the substrate. As $r$ decreases toward $r_*$, the tensile stress increases, and for $r < r_*$ material points undergo stick–slip motion.

We consider the force balance in an infinitesimal domain $d\Omega \equiv [r, r+dr] \times [\theta, \theta+d\theta]$, where $\times$ denotes the Cartesian product:

$$\frac{1}{r}\frac{\partial}{\partial r}(rh\sigma\hat{\boldsymbol{e}}_r) + \frac{1}{r}\frac{\partial}{\partial\theta}(h\sigma\hat{\boldsymbol{e}}_\theta) + \boldsymbol{f} = \boldsymbol{0}, \tag{S33}$$

where the stress tensor is expressed as $\sigma = \sigma_{\alpha\beta}\hat{\boldsymbol{e}}_\alpha \otimes \hat{\boldsymbol{e}}_\beta$, with $\otimes$ indicating the tensor product. For axisymmetric states, this equation reduces to the classical Lamé-type equilibrium equation [4]:

$$\frac{\sigma_{rr} - \sigma_{\theta\theta}}{r} + \frac{d}{dr}\sigma_{rr} + \frac{f_r}{h} = 0. \tag{S34}$$

We consider the case in which the boundary $r = r_*$ lies within the infinitesimal domain $d\Omega$. Using the maximum static friction at this boundary,

$$f_r(r_*) = \mu\rho gh, \tag{S35}$$

together with the overall force balance, we estimate the relationship between $d$ and $r_*$ as follows. In the planar region, the stresses decay as $\sigma_{rr}(r) \sim \sigma_{rr}(R)R^2/r^2$ and $\sigma_{\theta\theta}(r) \sim -\sigma_{rr}(R)R^2/r^2$. Using $\sigma_{rr}(R) \sim E(d/R)^2$ and Eq. (S34), and balancing the terms in Eq. (S34) at $r \sim r_*$, we obtain the scaling relation

$$r_* \sim \left(\frac{E}{\mu\rho g}\right)^{1/3}d^{2/3}. \tag{S36}$$

This scaling suggests that, in the absence of friction ($\mu = 0$), the stick–slip boundary $r_*$ diverges, so that the entire contact region slips for arbitrarily small indentation, consistent with physical intuition. Furthermore, by setting $r_* = a$ in Eq. (S36), we can estimate the critical indentation height $d_*$ at which the entire contact region begins to slide:

$$d_* \sim \left(\frac{\mu\rho g}{E}\right)^{1/2}a^{3/2}. \tag{S37}$$

We compare Eq. (S36) with the numerical results. In our finite-element simulations (FES), we define "substantial slip" by the criterion $|u_r(r)| > \varepsilon\Delta x$, where $\Delta x$ is the typical mesh size and $\varepsilon = 10^{-3}$ (See also §V in the main text). Using the parameter set as in Table II of the main text, we examined the azimuthally averaged radial displacement $\overline{u_r}(r) \equiv \frac{1}{2\pi}\oint u_r(r,\theta)d\theta$ over the entire contact region $R < r < a$. We then identified the location $r_*$ separating regions

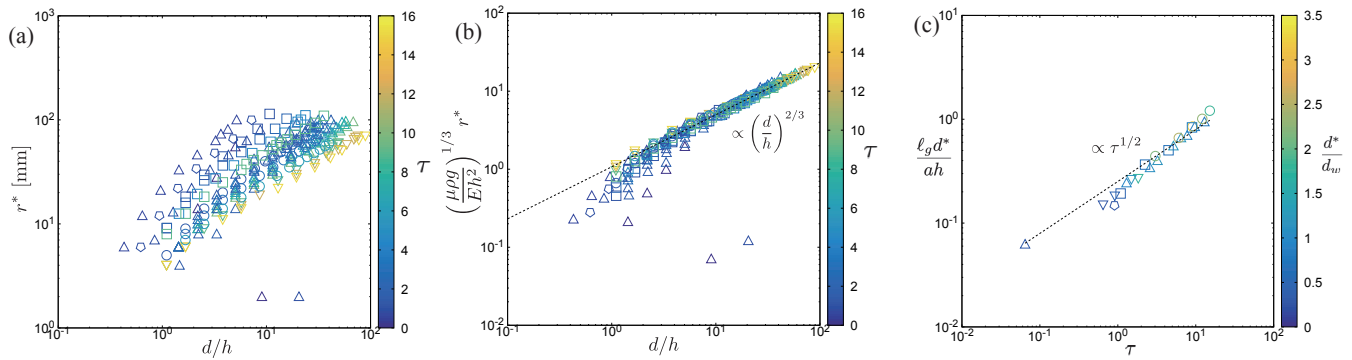

FIG. S2. Finite element simulation data on frictional slip in the contact region. (a) Relationship between the stick-slide boundary $r_*$ and indentation height $d$. The color bar indicates the frictional strength $\tau$. (b) Same data as in (a) but rescaled according to Eq. (S38); the dashed line represents Eq. (S38). (c) Critical indentation height $d_*$ at which the stick–slip boundary reaches the outer edge ($r_* = a$), as a function of $\tau$. The dashed line shows Eq. (S39). The color bar represents the ratio $d_*/d_w$, where $d_w$ is the critical displacement at the onset of wrinkling.

with substantial slip from those where the displacement remains below the threshold, $|u_r(r)| < \varepsilon \Delta x$. The results are shown in Figs. S2 (a) and (b), where the dashed line represents

$$\left(\frac{\mu \rho g}{Eh^2}\right)^{1/3} r_* \approx 1.07 \left(\frac{d}{h}\right)^{2/3},$$ (S38)

with the prefactor 1.07 determined from a fit to the FES results.

We also measured the critical displacement $d_*$ at which the outer edge of the sheet ($r = a$) begins to slip in our FES. Equation (S37) can be rewritten in terms of the relative frictional stress $\tau$ (see Eq. (32) in the main text) as $\ell_g d_*/(ah) \sim \tau^{1/2}$. Our numerical results, shown in Fig. S2 (c), are consistent with this scaling. By fitting the numerical data, we obtain the best-fit prefactor

$$\frac{\ell_g d_*}{ah} \approx 0.23 \ \tau^{1/2}.$$ (S39)

The color bar in Fig. S2 (c) represents the ratio $d_*/d_w$ measured in these FESs. For $\tau \lesssim 1$, $d_*/d_w$ tends to less than unity. This trend is consistent with the assumption used in the main text that "complete slip" occurs prior to the onset of wrinkling.

## IV. PLANE STRESS UNDER THE QUASI-STATIC SLIDING

Here, we derive the stress field in the contact region after complete slipping has occurred. This analysis allows us to clarify how friction affects the stress distribution relevant to the onset of wrinkling (see § VI in the main text).

As in the main text, we focus on the quasi-static indentation and idealize the sheet–substrate interface as being everywhere close to the Coulomb threshold, i.e. $f_t = \mu f_n$. In this effective description, the frictional force is taken to have an approximately uniform magnitude $f_r \approx \mu \rho g h$ over the entire contact region. Under this assumption, we solve the following differential equation:

$$\frac{\sigma_{rr} - \sigma_{\theta\theta}}{r} + \frac{d}{dr}\sigma_{rr} + \mu \rho g = 0.$$ (S40)

The sheet edge at $r = a$ is stress-free, while a finite tension $T$ is applied at $r = R$. Accordingly, the boundary conditions are

$$\sigma_{rr}(R) = T,$$ (S41)

$$\sigma_{rr}(a) = 0.$$ (S42)

Hereafter, we explicitly denote the frictional dependence of the stress by writing $\sigma_{\alpha\beta}(r; \mu)$; for example, $\sigma_{\alpha\beta}(r; 0)$ corresponds to the frictionless case.

We first consider the case of $\mu = 0$. The solution is

$$\sigma_{rr}(r; 0) = \frac{T}{1 - (R/a)^2} \left\{ \left(\frac{R}{r}\right)^2 - \left(\frac{R}{a}\right)^2 \right\} \sim T \frac{R^2}{r^2}, \tag{S43}$$

$$\sigma_{\theta\theta}(r; 0) = -\frac{T}{1 - (R/a)^2} \left\{ \left(\frac{R}{r}\right)^2 + \left(\frac{R}{a}\right)^2 \right\} \sim -T \frac{R^2}{r^2}. \tag{S44}$$

Here, the symbol "$\sim$" indicates an approximation valid near $r \sim R$ under the assumption that the sheet radius is sufficiently large, i.e., $R \ll a$.

For $\mu > 0$, additional terms proportional to the friction coefficient $\mu$ are superimposed on $\sigma_{\alpha\beta}(r; 0)$, as follows:

$$\sigma_{rr}(r; \mu) = \sigma_{rr}(r; 0) + \frac{2 + \nu}{3} \mu \rho g a \left\{ 1 - \frac{r}{a} - \frac{(R/r)^2 - (R/a)^2}{1 + R/a} \right\}, \tag{S45}$$

$$\sigma_{\theta\theta}(r; \mu) = \sigma_{\theta\theta}(r; 0) + \frac{2 + \nu}{3} \mu \rho g a \left\{ 1 - \frac{1 + 2\nu}{2 + \nu} \frac{r}{a} + \frac{(R/r)^2 + (R/a)^2}{1 + R/a} \right\}. \tag{S46}$$

The onset of wrinkling is governed by the stresses near $r \sim R$. For a sufficiently large sheet, i.e., under the assumption $R \ll a$, the hoop stress at $r = R$ is obtained as

$$\sigma_{\theta\theta}(R; \mu) = -T \left\{ 1 + O((R/a)^2) \right\} + \frac{2 + \nu}{3} \mu \rho g a \left\{ 2 - \frac{3(1 + \nu)}{2 + \nu} \frac{R}{a} + O((R/a)^2) \right\}. \tag{S47}$$

The parameter $T$ should, in principle, be determined by the matching solution in the lifted region for $r < R$, obtained from the FvK equations, to the outer planar solution given by Eqs. (S45) and (S46). Such a matching problem is left for future work. Instead, in the following subsection we use FESs to assess how $T$ depends on the friction coefficient $\mu$.

## A. Frictional dependence of the boundary tension $T$

To examine the $\mu$-dependence of $T$ numerically, we analyzed the stress profiles in the planar region using FES and fitted them to Eq. (S45) treating $T$ as a fitting parameter. The results for models with identical material and geometric parameters but different friction coefficients, $\mu$, are shown in Fig. S3. Each fitting was performed for the stress field at the indentation height corresponding to $d \approx d_w^0$. As shown in Fig. S3(b), the fitted values of $T$ increase sublinearly with $\mu$ from the value at $\mu = 0$ (denoted by $T_0$). Since the purpose of this section is to provide an order-of-magnitude estimation based on Eq. (S47), a detailed theoretical analysis of the $\mu$-dependence of $T$ is left for future work. Combining these numerical results with Eq. (S45), we find that the radial stress $\sigma_{rr}(r)$ exhibits only a weak dependence on $\mu$ near $r \sim R$ for small $\tau$. By contrast, the hoop stress $\sigma_{\theta\theta}(r)$ contains a term proportional to $\mu \rho g a$ as shown in Eq. (S47). This result justifies the assumption made in deriving Eqs. (34) and (35) in the main text: that $\sigma_{rr}(R)$ depends only weakly on $\tau$, whereas $\sigma_{\theta\theta}(R)$ exhibits a first-order dependence on $\tau$, at least for small $\tau$.

---

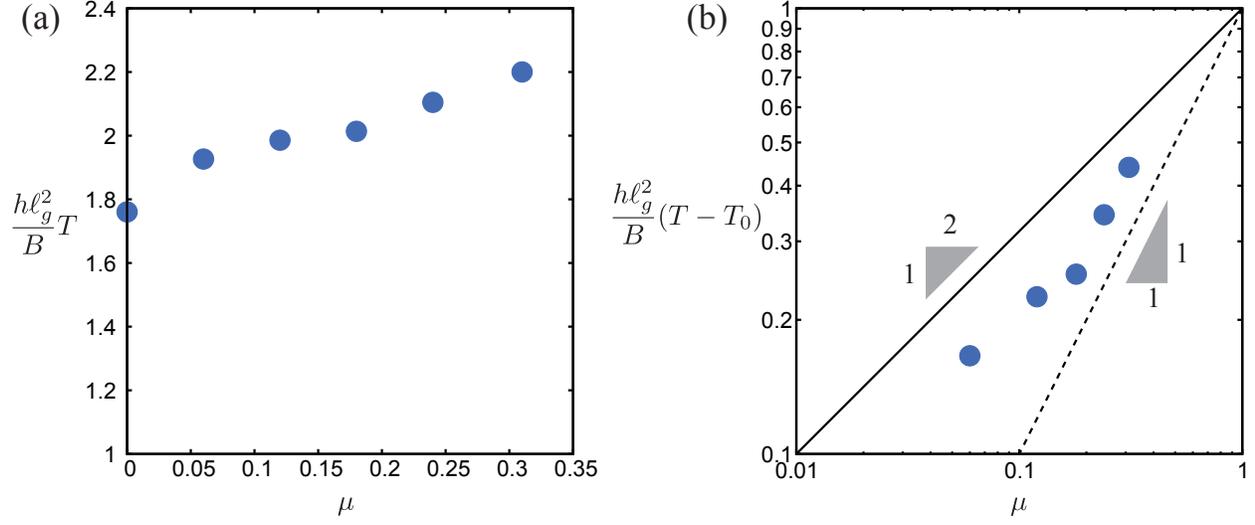

FIG. S3. Relationship between $T \equiv \sigma_{rr}(R)$ and the static friction coefficient $\mu$ obtained from FES. The elastic sheet parameters are $(E, \nu, \rho, h, a) = (477 \text{ kPa}, 0.47, 1127 \text{ kg/m}^3, 0.27 \text{ mm}, 100 \text{ mm})$. The value of $T$ was determined by fitting the azimuthally averaged radial stress $\overline{\sigma_{rr}}(r)$ at $d \approx d_w^0$ to Eqs. (S43) and (S45). (a) Dimensionless $T$ as a function of $\mu$. (b) The difference $T - T_0$ versus $\mu$, where $T_0$ denotes the value of $T$ at $\mu = 0$. The solid and dashed lines show reference curves proportional to $\mu^{1/2}$ and $\mu$, respectively.



\end{document}